\documentclass[%
 aip,
 amsmath,amssymb,
 reprint,%
author-year,%
]{revtex4-1}

\usepackage{graphicx}
\usepackage{dcolumn}
\usepackage{bm}
\usepackage{xcolor}
\usepackage[utf8]{inputenc}
\usepackage[T1]{fontenc}
\usepackage{etoolbox}
\usepackage{newtxtext}
\usepackage[varvw]{newtxmath}
\usepackage{threeparttable}
\usepackage{multirow}
\usepackage{enumerate}
\usepackage{enumitem}
\makeatletter
\def\@email#1#2{%
 \endgroup
 \patchcmd{\titleblock@produce}
 {\frontmatter@RRAPformat}
 {\frontmatter@RRAPformat{\produce@RRAP{*#1\href{mailto:#2}{#2}}}\frontmatter@RRAPformat}
 {}{}
}%
\makeatother

\begin{document}

\preprint{AIP/123-QED}

\title{Indirect X-ray photodesorption of $^{15}$N$_2$ and $^{13}$CO from mixed and layered ices}
\author{R. Basalgète}
\email{romain.basalgete@sorbonne-universite.fr}
\affiliation{Sorbonne Université, Observatoire de Paris, Université PSL, CNRS, LERMA, F-75005,
Paris, France}

\author{D. Torres-Díaz}%
\affiliation{Sorbonne Université, Observatoire de Paris, Université PSL, CNRS, LERMA, F-75005,
Paris, France}
\affiliation{Univ. Paris Saclay, CNRS, ISMO, 91405 Orsay, France}

\author{A. Lafosse}
 \affiliation{Univ. Paris Saclay, CNRS, ISMO, 91405 Orsay, France}

\author{L. Amiaud}
 \affiliation{Univ. Paris Saclay, CNRS, ISMO, 91405 Orsay, France}

\author{G. Féraud}
\affiliation{Sorbonne Université, Observatoire de Paris, Université PSL, CNRS, LERMA, F-75005,
Paris, France}

\author{P. Jeseck}
\affiliation{Sorbonne Université, Observatoire de Paris, Université PSL, CNRS, LERMA, F-75005,
Paris, France}

\author{L. Philippe}
\affiliation{Sorbonne Université, Observatoire de Paris, Université PSL, CNRS, LERMA, F-75005,
Paris, France}

\author{X. Michaut}
\affiliation{Sorbonne Université, Observatoire de Paris, Université PSL, CNRS, LERMA, F-75005,
Paris, France}

\author{J-H. Fillion}
\affiliation{Sorbonne Université, Observatoire de Paris, Université PSL, CNRS, LERMA, F-75005,
Paris, France}

\author{M. Bertin}
\affiliation{Sorbonne Université, Observatoire de Paris, Université PSL, CNRS, LERMA, F-75005,
Paris, France}

\date{\today}

\begin{abstract}
X-ray photodesorption yields of $^{15}$N$_2$ and $^{13}$CO are derived as a function of the incident photon energy near the N ($\sim$ 400 eV) and O K-edge ($\sim$ 500 eV) for pure $^{15}$N$_2$ ice and mixed $^{13}$CO:$^{15}$N$_2$ ices. The photodesorption spectra from the mixed ices reveal an indirect desorption mechanism for which the desorption of $^{15}$N$_2$ and $^{13}$CO is triggered by the photo-absorption of respectively $^{13}$CO and $^{15}$N$_2$. This mechanism is confirmed by the X-ray photodesorption of $^{13}$CO from a layered $^{13}$CO/$^{15}$N$_2$ ice irradiated at 401 eV, on the N 1s $\rightarrow \pi^*$ transition of $^{15}$N$_2$. This latter experiment enables to quantify the relevant depth involved in the indirect desorption process, which is found to be 30 - 40 ML in that case. This value is further related to the energy transport of Auger electrons emitted from the photo-absorbing $^{15}$N$_2$ molecules that scatter towards the ice surface, inducing the desorption of $^{13}$CO. The photodesorption yields corrected from the energy that can participate to the desorption process (expressed in molecules desorbed by eV deposited) do not depend on the photon energy hence neither on the photo-absorbing molecule nor on its state after Auger decay. This demonstrates that X-ray induced electron stimulated desorption (XESD), mediated by Auger scattering, is the dominant process explaining the desorption of $^{15}$N$_2$ and $^{13}$CO from the ices studied in this work.   
\end{abstract}

\maketitle

\section{\label{sec:intro}Introduction}
Diatomic molecules such as CO and N$_2$ have been extensively studied in laboratory at very low temperature (T < 20 K), in their solid phase. Indeed, these molecules are expected to play an important role in astrophysical environments. For instance, they are observed directly (especially for CO) or indirectly (especially for N$_2$) in the interstellar medium (ISM) and the solar system in a diversity of icy solids such as interstellar ices \citep{boogert_observations_2015}, comets \citep{Mumma_2011, Rubin_2015} and icy moons and planets \citep{Owen_1993, Cruikshank_1993, Bennet_2013}. Interaction of these icy solids with photons coming from nearby or distant stars can be at the origin of exchanges between solid and gas phases via the so-called photodesorption phenomenon. Photodesorption is a non-thermal process through which the energy of incident photons is absorbed by the ice and partly converted to the desorption of molecules from its surface. In the last decades, it has been extensively studied in the vacuum ultraviolet (VUV) range (7 - 13.6 eV) for N$_2$ and CO ices \citep{_berg_2007, oberg_photodesorption_2009, munoz_caro_new_2010, fayolle_co_2011, fayolle_wavelength-dependent_2013, bertin_indirect_2013}. 

More recently, the desorption induced by the irradiation of X-rays, referred to as X-ray photodesorption, was also highlighted as a significant process for ices composed of simple \citep{dupuy_x-ray_2018, dupuy_co_2021} and complex molecules \citep{jimenez-escobar_x-ray_2018, ciaravella_x-ray_2020, basalgete_complex_2021-1}. More importantly, it was suggested that X-ray photodesorption of neutral molecules was strongly dependent on the ice composition via indirect desorption mechanisms \citep{basalgete_complex_2021}. A detailed comprehension of these indirect processes on a microscopic scale is of primary importance to understand their implication in astrophysical environments. For a proper modeling of such processes, it is also required to access quantitative data concerning desorption yields. 

The X-ray photodesorption from condensed N$_2$ and CO has been previously studied for thin layers deposited on top of transition and noble metals \citep{frigo_observation_1998, feulner_recent_2000, romberg_atom-selective_2000, feulner_core-excitation-induced_2002}. In that case, desorption of neutral and ionic molecules (N$_2$, CO, CO$^+$) as well as fragments (N, N$^+$, N$^{2+}$, C, O, O$^+$, C$^+$) was found to strongly depend on singly or multiply valence-excited states, after decay of the core hole state. These states were associated with specific desorption mechanisms. In some cases, ultra-fast processes occurring during the core hole state lifetime were also suggested to play a role for the desorption of ionic fragments \citep{frigo_observation_1998}. 

For thick molecular ices, characterized by a higher number of layers deposited onto a substrate (generally more than 50 layers), the processes leading to desorption may be different. For instance, after X-ray absorption by molecules in the ice, the relaxation of the core hole state results in the emission of an Auger electron with a probability close to 1 for low Z elements \citep{Walters_1971, Krause_1979}. This Auger electron carries most of the initial photon energy and scatters inelastically in the ice, creating secondary events (e.g. ionization and excitation of the surrounding molecules) and especially, a cascade of secondary low energy (< 20 eV) electrons, that may lead in fine to the desorption of molecules from the ice surface. This process is known as X-ray induced electron stimulated desorption (XESD). XESD was not suggested to be the dominant process for the X-ray photodesorption of ions from pure N$_2$ and CO ice \citep{feulner_high_1992, dupuy_co_2021}. Whereas for the desorption of neutrals, it was assumed to be the case for pure CO ice \citep{dupuy_co_2021}, pure H$_2$O ice \citep{dupuy_desorption_2020} and pure and mixed methanol-containing ices \citep{basalgete_complex_2021-1, basalgete_complex_2021}. 

Based on quantitative data, X-ray photodesorption of ions from molecular ices is expected to be negligible compared to the desorption of neutrals \citep{dupuy_x-ray_2018, dupuy_co_2021}. In the present work, we experimentally study the X-ray photodesorption process for neutral desorption from pure $^{15}$N$_2$ ice and from mixed and layered $^{13}$CO:$^{15}$N$_2$ ices (isotopes are chosen to differentiate the molecules by mass spectrometry). Resonant 1s core excitation (or ionization) near the N ($\sim$ 400 eV) or O ($\sim$ 500 eV) K-edge, can be achieved thanks to tunable and high spectral resolution X-rays from synchrotron facilities, allowing thereby to selectively photo-excite $^{15}$N$_2$ or $^{13}$CO to trigger desorption from ices containing these molecules. We first provide a quantitative study of the X-ray photodesorption of neutral molecules from pure $^{15}$N$_2$ ice. Then, we quantify this process for mixed $^{13}$CO:$^{15}$N$_2$ and layered $^{13}$CO/$^{15}$N$_2$ ($^{13}$CO on top of $^{15}$N$_2$) ices. This latter set of experiments brings important information to discuss the desorption mechanisms. 

\section{\label{sec:methodo}Experiment and methodology}
\subsection{\label{sec:set-up}Setup and X-ray beamline}
To run the experiments, the Surface Processes and ICES (SPICES) setup was connected to the SEXTANTS beamline of the SOLEIL synchrotron facility at Saint-Aubin, France \citep{Sacchi_2013}. SPICES consists of an Ultra-High Vacuum (UHV) chamber (base pressure of $\sim$ 10$^{-10}$ mbar) at the center of which a copper substrate (polycrystalline oxygen-free high-conductivity copper, surface of 1.5 $\times$ 1.5 cm$^2$) can be cooled down to $\sim$ 15 K by a closed-cycle Helium cryostat. The substrate is electrically insulated from its sample holder by a Kapton foil. This enables the measurement of the drain current generated by electrons escaping the ice after X-ray absorption, which we referred to as the Total Electron Yield (TEY) in the following. The TEY can be measured while scanning the energy of the incident photons and it is representative of the X-ray absorption profile of the ices studied in this work. It is expressed in electrons per incident photon (displayed as e$^{-}$ photon$^{-1}$ for more simplicity) by correcting the drain current from the photon flux. 

The photon flux is measured by a calibrated silicon photodiode mounted on the beamline. The SEXTANTS beamline provides X-ray photons, energy of which can be finely tuned by means of a monochromator. The energy resolution can also be controlled by adjusting slits apertures at the exit of the monochromator. In the present studies, we have used photons in the 390 - 460 eV range (Nitrogen K-edge) and in the 530 - 560 eV range (Oxygen K-edge). Spectral resolutions of $\sim$ 100 meV and $\sim$ 350 meV were used to run the experiments. The photon flux varies with the energy and the spectral resolution between 10$^{11}$ and 10$^{12}$ photon s$^{-1}$. The beam was sent at a 47$^{\circ}$ incidence relative to the normal of the substrate surface and the spot area at the surface was approximately 0.1 cm$^{2}$. 

The calibration of the energy scale of the beamline was performed near the N K-edge and the O K-edge as followed: 
\begin{itemize}
 \item near the N K-edge, the TEY was measured on a pure $^{15}$N$_2$ ice as a function of the energy of the incident photons and it was compared to gas phase X-ray absorption spectroscopy experiments (e.g. \cite{chen_k_1989, feifel_quantitative_2004}). The TEY peak corresponding to the N 1s $\rightarrow$ $\pi^*$ ($v^{\prime}$ = 0) transition of $^{15}$N$_2$ was set to 400.868 eV according to \cite{chen_k_1989} (see Figure \ref{fig:TEY_N} and Section \ref{sec:TEY}). We therefore assumed that this energy position was not significantly shifted when going from gas phase to solid phase, as observed for a similar molecule: CO \citep{jugnet_high-resolution_1984}. Note that the energy of the transition is slightly varying ($\pm$ 20 meV) from one gas phase experiment to another \citep{KING_1977, HITCHCOCK_1980, SODHI_1984, Kempgens_1996, feifel_quantitative_2004, kato_absolute_2007}.
 \item near the O K-edge, the O 1s $\rightarrow$ $\pi^*$ transition of $^{13}$CO observed on the TEY of a mixed $^{13}$CO:$^{15}$N$_2$ ice was centered at 534.4 eV (see Figure \ref{fig:TEY_O} and Section \ref{sec:TEY}) according to pure CO ice spectroscopy experiments \citep{jugnet_high-resolution_1984}, assuming that the inter-molecular interactions between $^{13}$CO and $^{15}$N$_2$ are not significantly shifting the energy position of the electronic transition, which is reasonable considering such weakly interacting solids
\end{itemize}

\subsection{\label{sec:set-up}Ice deposition}
Within the SPICES setup, a tube connected to a gas introduction system, that can be translated a few millimeters in front of the substrate surface, enables to inject a partial pressure of gas-phase molecules that stick on the cold surface to form the molecular ices without significantly modifying the base pressure of the UHV chamber. The gaseous molecules used during this work were molecular nitrogen $^{15}$N$_2$ (98.6\% $^{15}$N purity, Eurisotop) and carbon monoxide $^{13}$CO (99\% $^{13}$C purity, Eurisotop). The ices thickness is expressed in monolayer (ML), equivalent to a molecule surface density of approximately 10$^{15}$ molecule cm$^{-2}$. Prior to the synchrotron experiments, the temperature-programmed desorption (TPD) technique was used to calibrate the rate and deposition time needed to grow a targeted ice thickness. The relative uncertainty on the number of ML deposited during a deposition procedure is about 10\% (see e.g. \cite{doronin_adsorption_2015}). 

The ices studied in this work were deposited at 15 K. The mixed $^{13}$CO:$^{15}$N$_2$ ices were grown by controlling in situ the partial pressure of each species injected in the UHV chamber independently (two different valves connected to the deposition tube were used). In this way, the mixing ratio is set by adjusting the partial pressures corresponding to each species. The experiment regarding the layered $^{13}$CO/$^{15}$N$_2$ ice (see Section \ref{sec:Layered}) was achieved by first depositing 50 ML of $^{15}$N$_2$ at 15 K. Then, on top of it, a given number of ML of $^{13}$CO was deposited and the ice was irradiated to measure the photodesorption yields. This operation (deposition of $^{13}$CO on top and irradiation) was sequentially repeated on the same ice until reaching $\sim$ 90 ML of $^{13}$CO on top of the $^{15}$N$_2$ ice. The number of ML of $^{13}$CO deposited at each step is displayed in Figure \ref{fig:Layered_exp} (see Section \ref{sec:Layered}).

\subsection{\label{sec:set-up}Derivation of the photodesorption yields}
The photodesorbed molecules are probed in the gas phase during the X-ray irradiation of the ice by means of a quadrupole mass spectrometer (QMS) equipped with an electron impact ionization chamber at 70 eV. The photodesorption of $^{13}$CO and $^{15}$N$_2$ was monitored by recording the QMS signal on the mass channel 29 u. and 30 u. respectively. The signal intensities are divided by the photon flux and multiplied by a factor $k$ to obtain the X-ray photodesorption yields $\Gamma_{raw}$ expressed in molecule desorbed per incident photon (displayed as molecules photon$^{-1}$ in the following for more simplicity). The factor $k$ was derived prior to the synchrotron runs to relate the QMS current to a calibrated number of $^{15}$N$_2$ molecules desorbed during a TPD experiment. For more details on the calibration procedure, see \cite{basalgete_complex_2021}. The apparatus function of our QMS can be considered constant between the mass channels 29 u. and 30 u. and the non-dissociative electron impact ionization cross sections at 70 eV of $^{13}$CO \citep{tian_cross_1998} and $^{15}$N$_2$ \citep{straub_absolute_1996} are nearly equals such that we considered that the detection efficiency of our QMS is similar for $^{13}$CO and $^{15}$N$_2$. This results in using a similar factor $k$ for quantifying the photodesorption yields of $^{13}$CO and $^{15}$N$_2$. The systematic uncertainty on the photodesorption yields provided by this calibration method is estimated to be $\sim$ 50\%.

In Section \ref{sec:comp_VUV}, we compute the photodesorption yields $\Gamma_{eV}$ expressed in molecule desorbed per eV deposited by correcting the experimental photodesorption yields $\Gamma_{raw}$ (expressed in molecules photon$^{-1}$) considering the absorption cross section. This results in the following formula:
\begin{equation}
\label{eq:gamma_ev}
\Gamma_{eV}(E) = \frac{\Gamma_{raw}(E)}{E \, a_{dil} \, (1 - e^{- \sigma(E) N})} 
\end{equation}
where $E$ is the energy of the incident photon, $a_{dil}$ is a factor taking into account the dilution of the photodesorbing molecules in the case of mixed ices (namely $a_{dil}$ = 1/3 and 2/3 for the photodesorption yield of, respectively, $^{15}$N$_2$ and $^{13}$CO from a mixed $^{13}$CO:$^{15}$N$_2$ - 2:1 ice), $\sigma(E)$ is the photo-absorption cross section at $E$ and $N$ is the column density involved in the desorption process. $N$ is equal to $N = \Lambda_{des} \times b_{dil} \times 10^{15}$ cm$^{-2}$ where $\Lambda_{des}$ is the desorption-relevant depth expressed in ML (this notion is explained in more details in Section \ref{sec:discussion}) and $b_{dil}$ is a factor taking into account the dilution of the photo-absorbing molecule in the case of mixed ices (namely $b_{dil}$ = 1/3 and 2/3 for the yields derived, respectively, near the N and O K edge). 
\begin{figure}[h!]
\includegraphics[width=8.5cm]{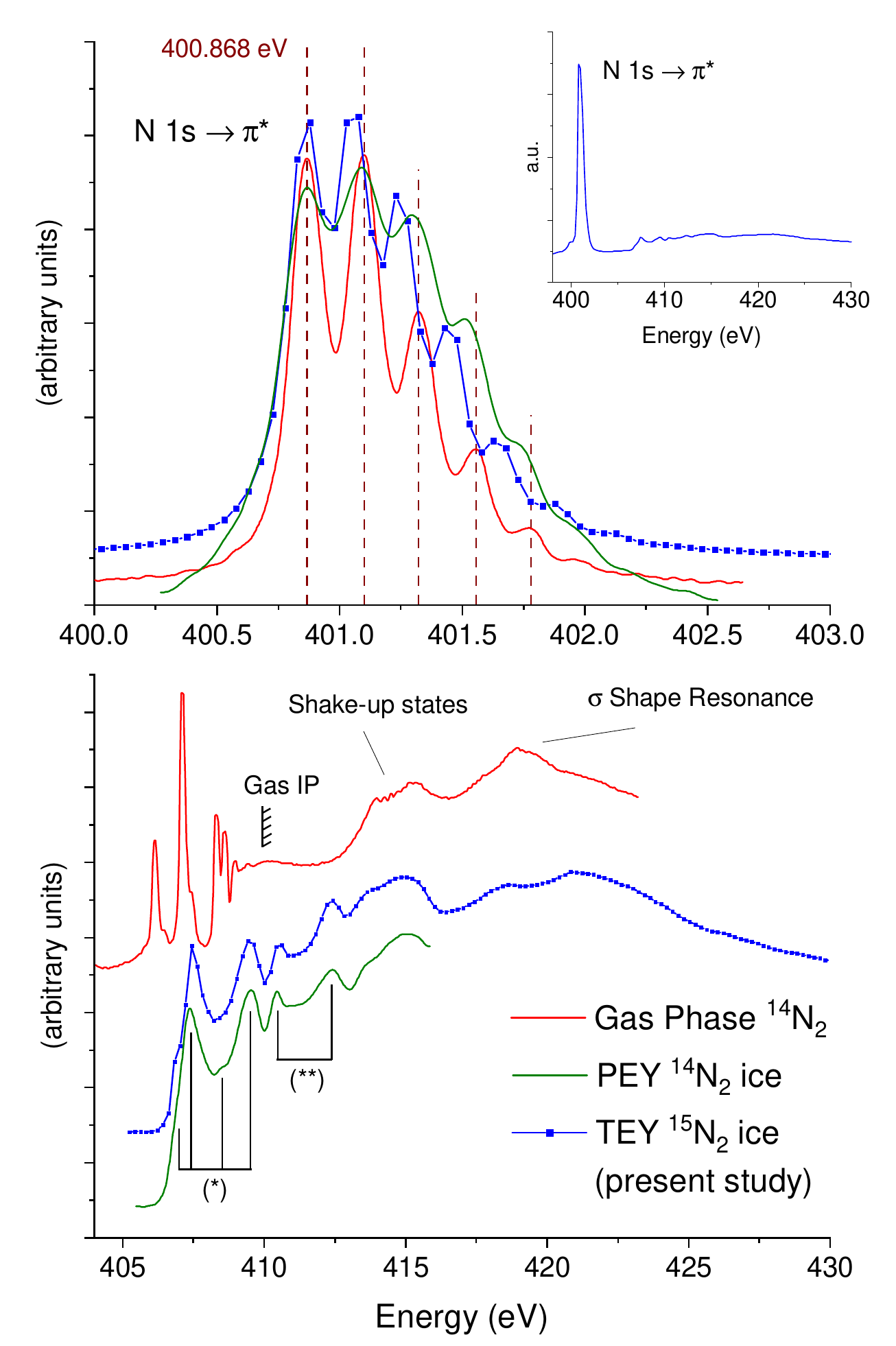}
\caption{\label{fig:TEY_N} Upper right panel: TEY of $^{15}$N$_2$ ice at 15 K (50ML) from our experiments in the N K-edge region between 398 and 430 eV. The upper left panel shows the N 1s $\rightarrow \pi^*$ excitation region and the lower panel shows the region near the ionization threshold between 405 and 430 eV. In solid red lines are the X-ray photoabsorption spectra of gas phase $^{14}$N$_2$ from \cite{chen_k_1989} with a spectral resolution of 40 meV. In solid green lines are the partial electron yields (PEYs) of 50 ML of $^{14}$N$_2$ ice at 15 K from \cite{feulner_high_1992} with a spectral resolution of 100 meV. In blue lines are the TEYs of 50 ML of $^{15}$N$_2$ ice at 15 K from our experiments with a spectral resolution of 100 meV for the upper left panel and 350 meV for the lower and upper right panels. The baselines of the different curves have been shifted for more clarity. The red dashed vertical lines in the upper left panel are centered on the gas phase vibrational bands.}
\end{figure}

\section{\label{sec:Results}Results}
\subsection{\label{sec:TEY}X-ray absorption spectra}
\subsubsection{$^{15}$N$_2$ ice photoabsorption near N K-edge}

The X-ray absorption spectrum of pure $^{15}$N$_2$ ice at 15 K near the N K-edge, assimilated to our TEY measurements, is displayed in Figure \ref{fig:TEY_N} (blue lines) for different energy ranges and spectral resolutions. The absorption is dominated by the N 1s $\rightarrow \pi^*$ transition of $^{15}$N$_2$ near 401 eV, whose vibrational structure has been resolved with a spectral resolution of 100 meV. Our TEY in the N 1s $\rightarrow \pi^*$ region (see upper panel of Figure \ref{fig:TEY_N}) compares well with other absorption spectra of gas phase $^{14}$N$_2$ \citep{chen_k_1989} and $^{14}$N$_2$ ice at 15 K \citep{feulner_high_1992}. The N 1s$^{-1} \pi^*$ core electronic state of the free molecule is not significantly modified in the solid phase, as expected for weak Van der Waals interactions between N$_2$ molecules in the solid. There are however some slight discrepancies in the position of the vibrational levels between the three curves in the upper panel of Figure \ref{fig:TEY_N}. When going from gas phase (red line) to solid phase (green line), the small differences in the position of the vibrational levels of the N 1s$^{-1}\pi^*$ state of $^{14}$N$_2$ could be attributed to either experimental uncertainties on the relative position of the peaks or to inter-molecular interactions in the solid phase. The comparison of our TEY of $^{15}$N$_2$ ice (blue line) with the PEY of $^{14}$N$_2$ ice (green line) shows a smaller vibrational level spacing for $^{15}$N$_2$. This is expected to come from an isotopic effect. A similar behaviour is observed in the vibrational level spacing of the A$^{1} \Pi$ state of $^{13}$CO versus $^{12}$CO ice and of the b$^1\Pi_u$ state of $^{15}$N$_2$ versus $^{14}$N$_2$ ice probed by Vacuum UltraViolet (VUV) photodesorption experiments \citep{bertin_uv_2012, fayolle_wavelength-dependent_2013}.
\begin{figure}[t!]
\includegraphics[width=8.5cm]{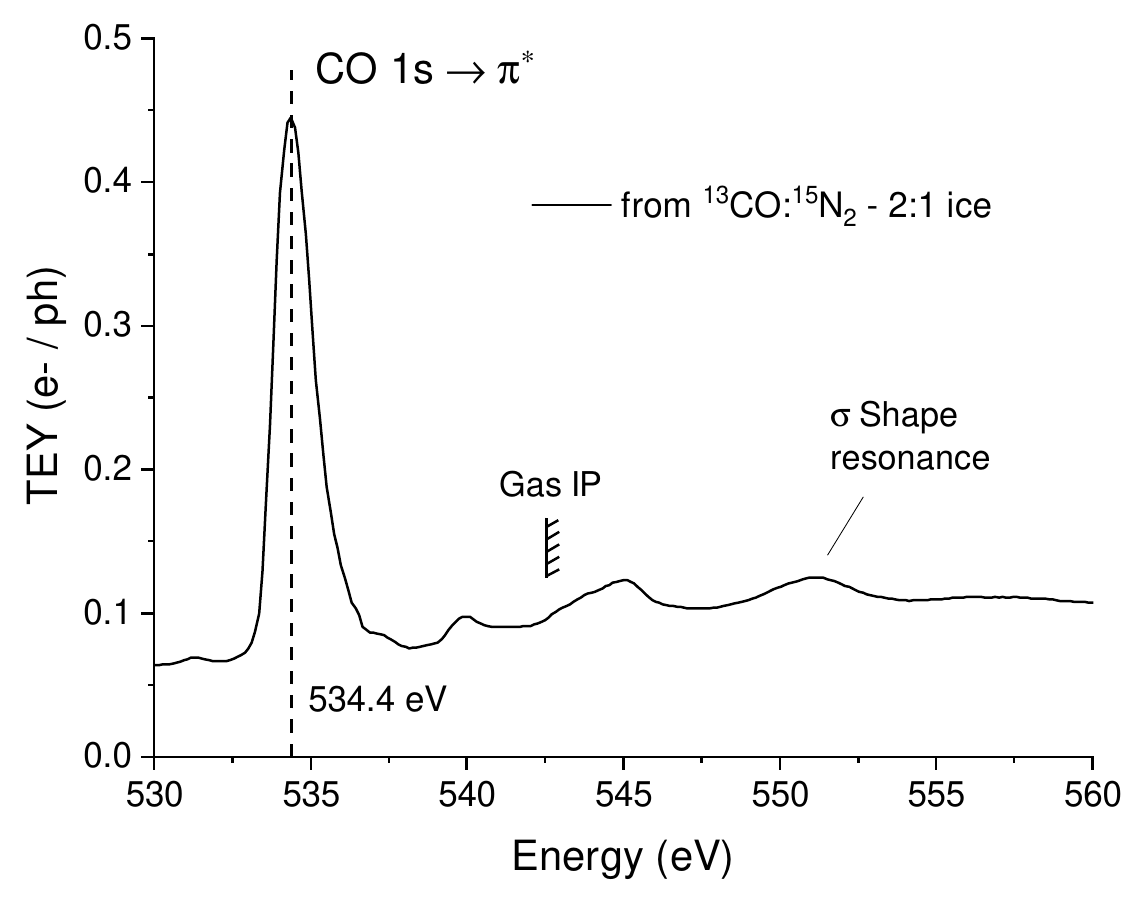}
\caption{\label{fig:TEY_O} TEY of a mixed $^{13}$CO:$^{15}$N$_2$ - 2:1 ice at 15 K (total of 100 ML) in the O K-edge region between 530 and 560 eV from our experiments, with a spectral resolution of 350 meV.}
\end{figure}
In the lower panel of Figure \ref{fig:TEY_N}, the Rydberg states of gas phase $^{14}$N$_2$ seen below the ionization potential (IP) are significantly modified in the solid phase, as expected from the overlapping of the Rydberg orbitals with that of the neighboring molecules in the solid. Instead, series of broad peaks (labelled (*) and (**) in Figure \ref{fig:TEY_N}), whose positions go above the gas phase IP, populate this region. There is no significant difference in this region between $^{14}$N$_2$ and $^{15}$N$_2$ in solid phase. The first series of the four peaks labelled (*) has been attributed to the equivalent of the gas phase Rydberg states 3s$\sigma$, 3p$\pi$, 4s$\sigma$ and 5s$\sigma$ in solid phase by \cite{feulner_high_1992}, based on polarization dependent X-ray photodesorption of N$^+$ and based on the data from \cite{chen_k_1989}. The attribution of the peaks labelled (**) above the gas phase IP in Figure \ref{fig:TEY_N} has not been studied in the literature to our knowledge. The gas phase broad feature observed near 415 eV has been attributed by \cite{chen_k_1989} to shake-up states, temptatively associated with the state configurations (1s)$^{-1}$(5$\sigma$)$^{-1}$(1$\pi_g^*$)$^1$(3s$\sigma$)$^1$ and (1s)$^{-1}$(1$\pi$)$^{-1}$(1$\pi_g^*$)$^1$(3p$\pi$)$^1$. As one can see in Figure \ref{fig:TEY_N}, a similar feature is observed in this region for $^{15}$N$_2$ and $^{14}$N$_2$ ice. The feature observed near 419 eV in the gas phase has been attributed to a $\sigma$ shape resonance \citep{chen_k_1989}. A similar feature is observed for $^{15}$N$_2$ ice.

\subsubsection{$^{13}$CO:$^{15}$N$_2$ ice photoabsorption near O K-edge}
The TEY of a mixed $^{13}$CO:$^{15}$N$_2$ ice (2:1) ice at 15 K near the O K-edge is displayed in Figure \ref{fig:TEY_O}. The absorption is dominated by the O 1s $\rightarrow \pi^*$ transition of $^{13}$CO centered at 534.4 eV. The spectrum is similar to that of a pure CO ice studied in similar experiments and for which more details are available regarding the attribution of the spectral features \citep{dupuy_co_2021}. The inter-molecular interactions between $^{13}$CO and $^{15}$N$_2$ in the mixed ice are not modifying the transitions observed in this energy range compared to a pure CO ice. The N 1s core electron photo-ionization of $^{15}$N$_2$ is expected to contribute to the absorption spectrum in this energy range but it is reasonably assumed to be negligible compared to $^{13}$CO photo-absorption, and it should in any case account for a monotonically decreasing signal over the whole energy range.
\begin{figure}
\includegraphics[width=8.5cm]{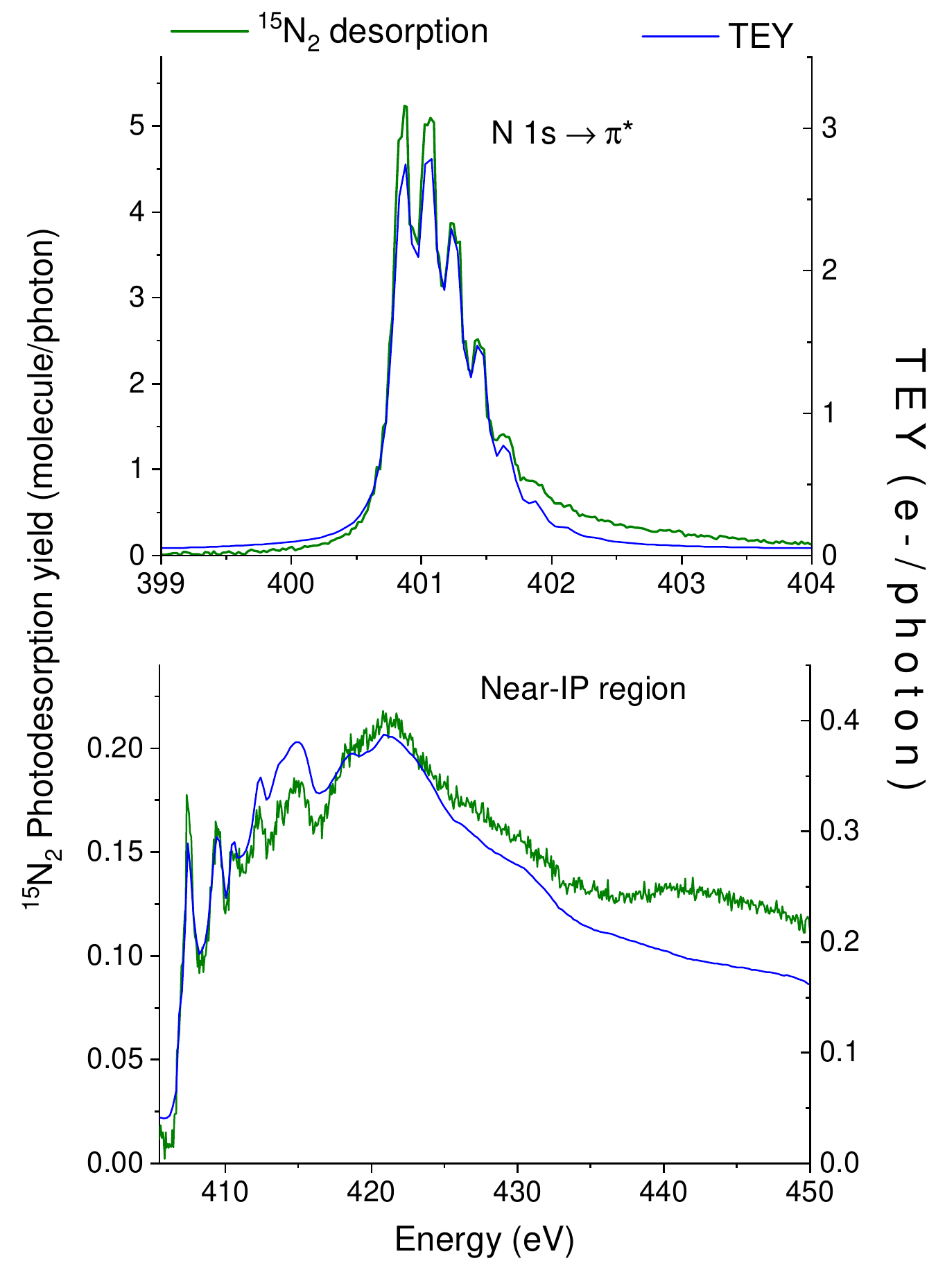}
\caption{\label{fig:Yield_N_pure} X-ray photodesorption yields of $^{15}$N$_2$ (green lines) from pure $^{15}$N$_2$ ice at 15 K (50 ML) in the N 1s $\rightarrow \pi^*$ excitation region (upper panel, spectral resolution of 100 meV) and in the near-IP region (lower panel, spectral resolution of 350 meV). The TEYs measured during the irradiations are also displayed for comparison (in blue lines).}
\end{figure}

\begin{figure*}
\includegraphics[width=14.6cm]{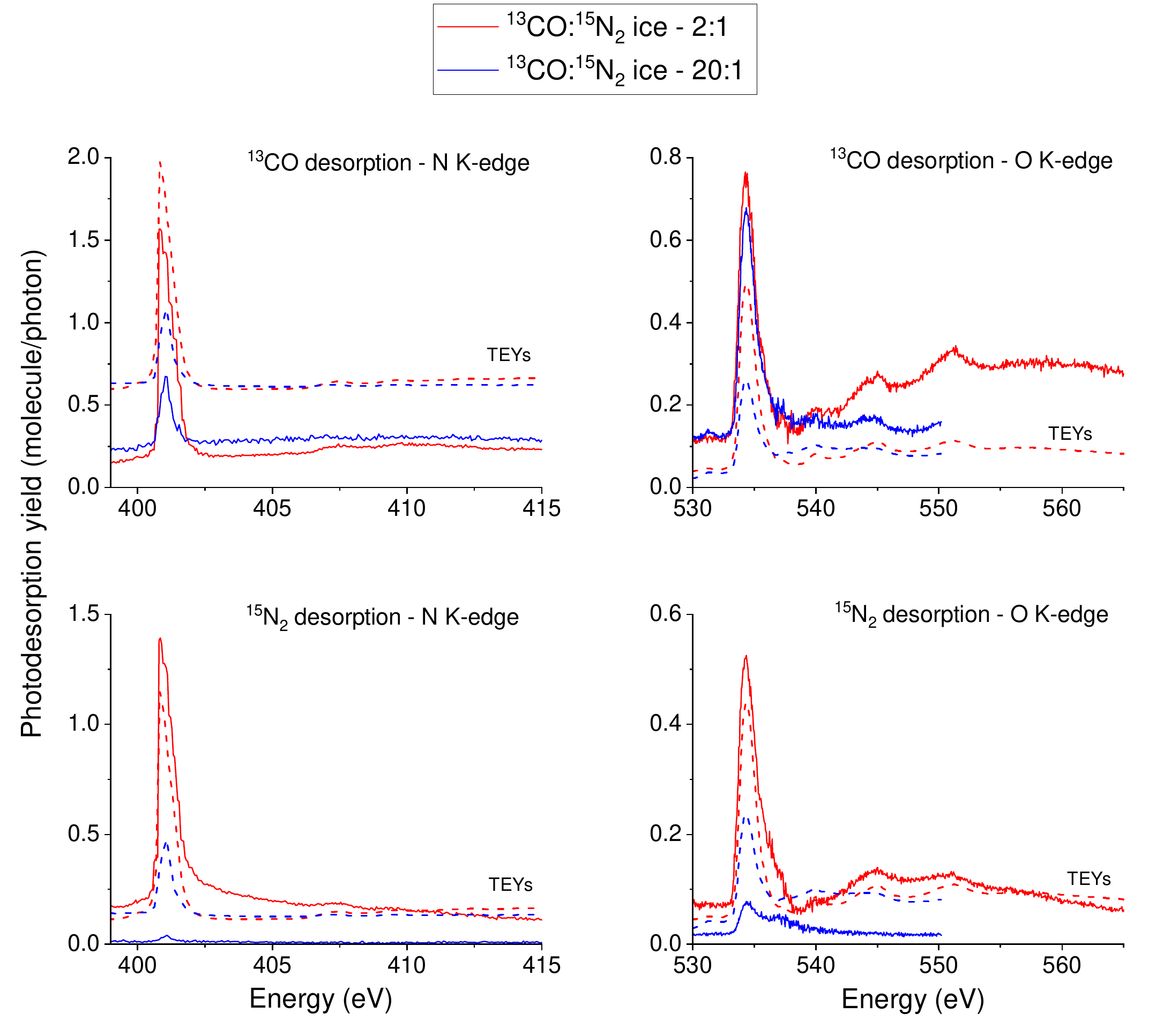}
\caption{\label{fig:Yield_Mix} X-ray photodesorption yields of $^{13}$CO and $^{15}$N$_2$ (solid lines) in the N K-edge (left panels) and O K-edge (right panels) regions from mixed $^{13}$CO:$^{15}$N$_2$ ices at 15 K (total of 100 ML) with mixing ratio of 2:1 (red lines) and 20:1 (blue lines). The TEYs measured simultaneously during the corresponding irradiation procedure are also displayed in dashed lines in arbitrary units.} 
\end{figure*}

\subsection{\label{sec:Yields}X-ray photodesorption yields}
\subsubsection{\label{sec:Yields_pure}$^{15}$N$_2$ pure ice}

The photodesorption spectra of $^{15}$N$_2$ from a pure $^{15}$N$_2$ ice at 15 K are shown in Figure \ref{fig:Yield_N_pure} (green lines) with the upper panel corresponding to the N 1s $\rightarrow \pi^*$ transition and the lower panel corresponding to the near-IP region. The TEYs measured simultaneously during the irradiation procedure are also displayed in blue lines for comparison. The photodesorption yields are higher by more than one order of magnitude in the N 1s $\rightarrow \pi^*$ region than in the near-IP region, with a maximum of $\sim$ 5 molecules photon$^{-1}$ at the N 1s $\rightarrow \pi^*$ ($v^{\prime}$ = 0, 1) transitions. The photodesorption spectra are following the same variations as the TEYs except for small discrepancies on the relative intensities of the features, that are likely due to background correction issues and for which we do not attribute any physical meaning. The photodesorption yields of $^{15}$N$_2$ and the TEYs are obtained from an ice having received a fluence $< 3.10^{16}$ ph cm$^{-2}$. They both decrease with increasing fluence due to the ice photo-ageing, that is the structural and chemical modifications of the ice due to photo-absorption, but the shape of the spectra remains the same and no significant new absorption features are observed in the energy range probed (400 - 450 eV). The possible photodesorption of $^{15}$N$_3$ and $^{15}$N$_4$ were monitored during the irradiation but no desorption signals were found (with a detection limit of 10$^{-2}$ molecules photon$^{-1}$ considering the mass channels and photon flux). Photodesorption of atomic $^{15}$N was only detectable (with a detection limit of 10$^{-3}$ molecules photon$^{-1}$) at the N 1s $\rightarrow \pi^*$ transition, with a yield of $\sim$ 0.01 molecules photon$^{-1}$, which is two orders of magnitude less than the desorption yield of $^{15}$N$_2$.  

\subsubsection{Mixed $^{13}$CO:$^{15}$N$_2$ ice}

The photodesorption yields of $^{13}$CO and $^{15}$N$_2$ from mixed $^{13}$CO:$^{15}$N$_2$ (2:1 and 20:1) ices are displayed in Figure \ref{fig:Yield_Mix} in the N K-edge and O K-edge energy range. The TEYs measured simultaneously during the irradiation procedures are also displayed in dashed lines and vertically shifted for more clarity. The shape of the photodesorption spectra are well-correlated to the TEYs both at the N and O K-edge, similarly to what is observed for a pure $^{15}$N$_2$ ice. There is however a small discrepancy: in the lower panels of Figure \ref{fig:Yield_Mix}, the photodesorption yield of $^{15}$N$_2$ from a mixed $^{13}$CO:$^{15}$N$_2$ (2:1) ice is decreasing after 403 eV for the N K-edge and after 555 eV for the O K-edge whereas the corresponding TEY is constant. This is due to background correction issues. 

The photodesorption of $^{15}$N$_2$ from mixed $^{13}$CO:$^{15}$N$_2$ ices is significantly decreasing between the 2:1 and 20:1 ratio both at the N and O K-edge (lower panels of Figure \ref{fig:Yield_Mix}) due to the high dilution of $^{15}$N$_2$ whereas $^{13}$CO desorption yield is in any case found in the same order of magnitude (upper panels). In the case of a mixed $^{13}$CO:$^{15}$N$_2$ (2:1) ice, the photodesorption yields of $^{13}$CO and $^{15}$N$_2$ are found in the same order of magnitude at the N 1s $\rightarrow \pi^*$ ($\sim 1.4 - 1.5$ molecules photon$^{-1}$) and the O 1s $\rightarrow \pi^*$ ($\sim 0.5 - 0.7$ molecules photon$^{-1}$) transitions of $^{15}$N$_2$ and $^{13}$CO respectively. This is consistent with the fact that CO and N$_2$ have similar binding energies in mixed CO:N$_2$ ices \citep{oberg_competition_2005, bisschop_desorption_2006}. The fact that the photodesorption yields are lower at the O 1s $\rightarrow \pi^*$ transition compared to the N 1s $\rightarrow \pi^*$ transition in the case of the 2:1 ratio is probably due to a higher absorption cross section (based on gas phase experiments) of $^{15}$N$_2$ at 401 eV ($\sim$ 26 Mbarn from \cite{kato_absolute_2007} and when considering the spectral resolution of our photon source) compared to that of $^{13}$CO at 534.4 eV ($\sim$ 2.8 - 3.2 Mbarn from \cite{barrus_k_1979}).

Despite a higher photoabsorption of $^{13}$CO expected for a dilution ratio of 20:1 compared to 2:1, the TEY value at the O 1s $\rightarrow \pi^*$ transition of $^{13}$CO is lower for the 20:1 ratio (see right panels of Figure \ref{fig:Yield_Mix}). We would have expected the photodesorption yield of $^{13}$CO to also follow this trend. However, the photodesorption yield of $^{13}$CO at this energy is not significantly changing from a 2:1 to a 20:1 ratio (the yield is at $\sim$ 0.7 molecules photon$^{-1}$). A higher fluence received by the $^{13}$CO:$^{15}$N$_2$ (20:1) ice before this specific measurement seems to be responsible for this behaviour. The fluences corresponding to the other measurements displayed in Figure \ref{fig:Yield_Mix} are similar (< 2 $\times$ 10$^{16}$ ph cm$^{-2}$).

Photodesorption signals were observed on the mass channels 28 u. and 44 u.. They could correspond to the desorption of $^{13}$C$^{15}$N and $^{13}$C$^{15}$NO respectively. However, the signals were too weak (S/N $\gtrsim$ 1, with a photodesorption yield of $\sim$ 0.02 molecules photon$^{-1}$ at the N 1s $\rightarrow \pi^*$ and the O 1s $\rightarrow \pi^*$ for both mass channels) and more easily detected in the case of a 20:1 ratio such that we cannot totally exclude the contribution of $^{12}$CO, $^{14}$N$_2$ (mass 28 u.) and $^{12}$CO$_2$ (mass 44 u.) photodesorption to these mass channels due to natural isotopes in our gas samples ($\sim$ 1\%). Finally, a desorption signal was found on the mass channel 31 u., which is associated with the X-ray photodesorption of $^{15}$NO. The photodesorption spectra of $^{15}$NO are also following the same variations as the TEYs and the corresponding yields are $5.3 \times 10^{-2}$ molecules photon$^{-1}$ at the N 1s $\rightarrow \pi^*$ transition of $^{15}$N$_2$ and $2.7 \times 10^{-2}$ molecules photon$^{-1}$ at the O 1s $\rightarrow \pi^*$ transition of $^{13}$CO for a $^{13}$CO:$^{15}$N$_2$ (2:1) ice. The formation of $^{15}$NO is expected to originate from chemistry induced by the dissociation of $^{15}$N$_2$ and $^{13}$CO.  Its presence in the ice has also been evidenced, in very small quantities, by TPD of the photo-irradiated ice at the end of the experiment. The photodesorption yields of the previous photo-products (possibly $^{13}$C$^{15}$N and $^{13}$C$^{15}$NO for the mass 28 u. and 44 u. respectively, and unambiguously $^{15}$NO for the mass 31 u.) are at least an order of magnitude lower than the yields of $^{13}$CO and $^{15}$N$_2$. This shows that, if some photo-chemistry occurs, its impact on the photodesorption process from mixed ices is limited. Photodesorption of atomic $^{15}$N, $^{13}$C and O was not detected (with a detection limit of 10$^{-3}$ molecules photon$^{-1}$). 

\subsubsection{\label{sec:Layered}Layered $^{13}$CO/$^{15}$N$_2$ ice}

\begin{figure}
\includegraphics[width=8.5cm]{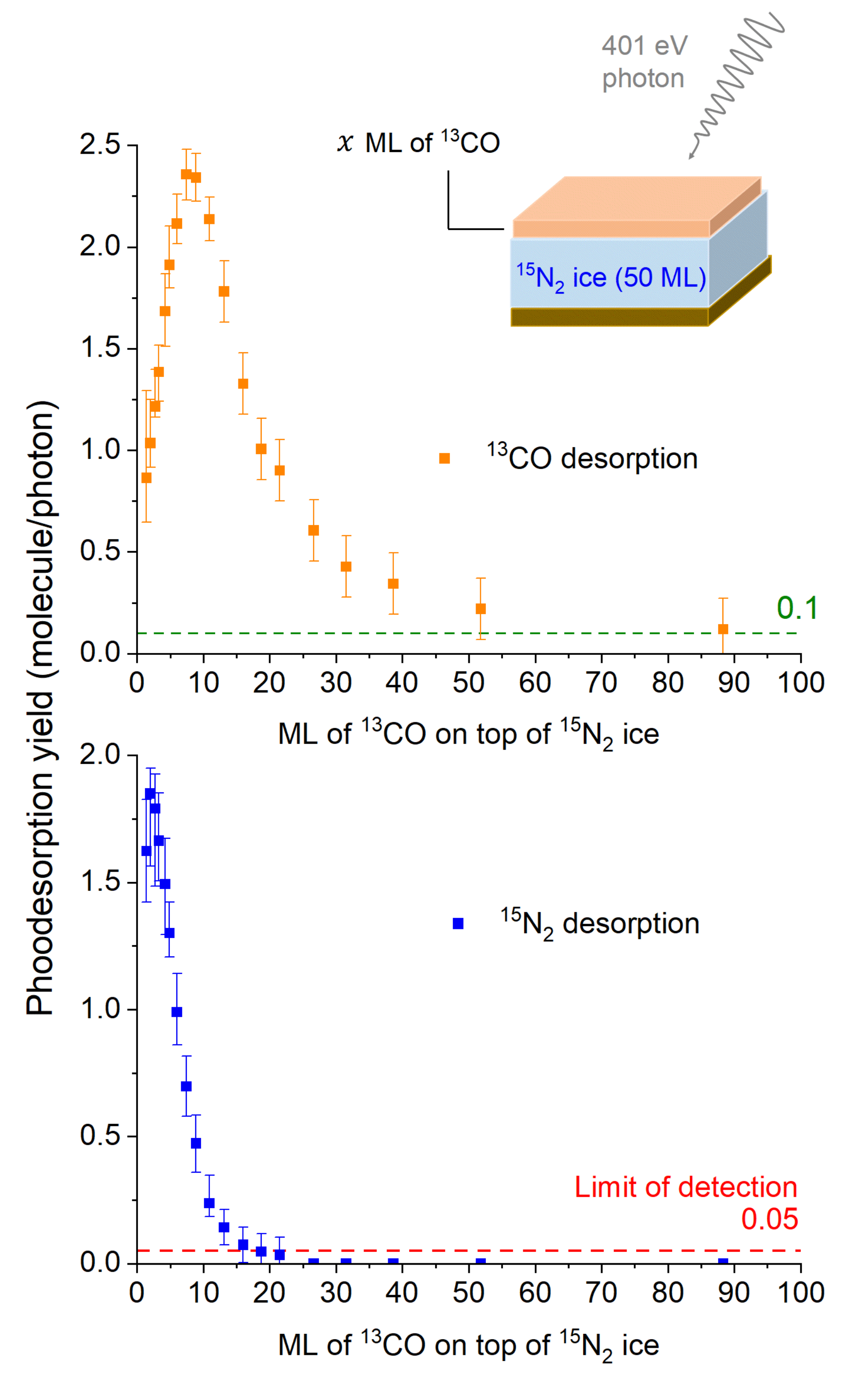}
\caption{\label{fig:Layered_exp} X-ray photodesorption yields at 401 eV of $^{13}$CO (upper panel) and $^{15}$N$_2$ (lower panel) from a layered $^{13}$CO / $^{15}$N$_2$ ice at 15 K, as a function of the number of ML of $^{13}$CO deposited on top of a 50 ML $^{15}$N$_2$ ice.}
\end{figure}

The photodesorption yields at 401 eV of $^{13}$CO and $^{15}$N$_2$ from a layered $^{13}$CO/$^{15}$N$_2$ ice are displayed in Figure \ref{fig:Layered_exp} as a function of the number of ML of $^{13}$CO on top of a $^{15}$N$_2$ ice. For this experiment, the photon flux was low ($\sim 10^{11}$ photon s$^{-1}$) and the irradiation time was short (a few seconds) to ensure that the number of $^{13}$CO molecules photodesorbed during one irradiation step are negligible compared to the number of ML of $^{13}$CO deposited on top of the $^{15}$N$_2$ ice. Considering a X-ray photodesorption yield of $^{13}$CO between 1.0 and 2.4 molecules photon$^{-1}$, an equivalent of 0.01 - 0.03 ML of $^{13}$CO is desorbed during one irradiation step. In comparison, the first point measured in Figure \ref{fig:Layered_exp} is at 1.4 ML of $^{13}$CO on top of the $^{15}$N$_2$ ice. We do not expect significant structural changes of the ice during each irradiation step at such low fluence. Chemical changes are also not expected based on the fact that no signs of significant chemistry are observed on both the TEYs and in desorption for the previous pure and mixed irradiated ices. 

At 401 eV, photo-absorption of $^{15}$N$_2$ is due to the N 1s $\rightarrow \pi^*$ transition and photo-absorption of $^{13}$CO is due to the ionization of its C 1s electron (IP $\sim$ 296 eV). Quantitatively speaking, and based on gas phase data, the absorption cross section of $^{15}$N$_2$ and $^{13}$CO at 401 eV are $\sim$ 20 - 40 Mbarn and $\sim$ 0.3 Mbarn respectively \citep{barrus_k_1979, kato_absolute_2007}. Hence, photo-absorption at 401 eV in our layered $^{13}$CO/$^{15}$N$_2$ ice is dominated by $^{15}$N$_2$ photo-absorption. The idea of this experiment is therefore to trigger the desorption of the above lying $^{13}$CO molecules by photo-excitation of the supporting $^{15}$N$_2$ molecules and to estimate the depth of this indirect desorption mechanism (see Section \ref{sec:discussion}) by sequentially depositing monolayers of $^{13}$CO on top of the ice and irradiate it. The contribution of $^{13}$CO photo-absorption to the desorption of $^{13}$CO at 401 eV can be estimated by considering that it is quantitatively similar to the photodesorption of CO from a pure CO ice at 530 eV, which exhibits a photodesorption yield of $\sim$ 0.1 molecules photon$^{-1}$ \citep{dupuy_co_2021}. This stems from the fact that the photo-absorption of $^{13}$CO at 530 eV is also due to the ionization of its C 1s electron and that we do not expect the ionization cross section to significantly differs between 401 to 530 eV. Consequently, the photodesorption yield of 0.1 molecules photon$^{-1}$ is used as a reference to estimate the contribution of $^{13}$CO photo-absorption to the desorption of $^{13}$CO at 401 eV in our layered ice experiment (it is displayed as a horizontal dashed green line in the upper panel of Figure \ref{fig:Layered_exp}). From the upper panel of Figure \ref{fig:Layered_exp}, we observe two different regimes for the photodesorption yield of $^{13}$CO:
\begin{itemize}
 \item a low to medium coverage regime, for a number of ML of $^{13}$CO from 1 to roughly 30 - 40 ML. In this regime, the photodesorption yield of $^{13}$CO is $>$ 0.1 - 0.2 molecules photon$^{-1}$ such that the desorption of $^{13}$CO is dominantly triggered by photo-excitation of $^{15}$N$_2$. In that case, the photodesorption yield of $^{13}$CO is increasing until $\sim$ 10 ML and then starts to decrease
 \item a high coverage regime, for a number of ML of $^{13}$CO $>$ 40 ML. In this regime, the photodesorption yield of $^{13}$CO is close to 0.1 molecules photon$^{-1}$ such that the desorption of $^{13}$CO is dominantly triggered by photo-ionization of $^{13}$CO. In that case, the possible photodesorption of $^{13}$CO triggered by photo-excitation of $^{15}$N$_2$ can be neglected
\end{itemize}

The photodesorption yield of $^{15}$N$_2$ from our layered ice is displayed in the lower panel of Figure \ref{fig:Layered_exp}. It is decreasing as the number of ML of $^{13}$CO increases until it goes below our detection limit of $\sim$ 0.05 molecules photon$^{-1}$ (considering the low photon flux for this experiment) for a number of ML of $^{13}$CO $>$ 15 ML. The fact that we still observe $^{15}$N$_2$ desorption even when the $^{15}$N$_2$ ice is covered by a few ML of $^{13}$CO may be due to diffusion of $^{15}$N$_2$ from below the ML of $^{13}$CO or to a few $^{15}$N$_2$ molecules still present at the surface due to inhomogeneous deposition of $^{13}$CO. This is further discussed in Section \ref{sec:model}.

\section{\label{sec:discussion}Discussion}
\subsection{\label{sec:indirect} Indirect desorption and estimation of $\Lambda_{des}$}

Our experiments on mixed $^{13}$CO:$^{15}$N$_2$ ices highlight an indirect desorption mechanism, in the sense that the photo-absorption of one molecule is inducing the desorption of another one. This is clearly seen in the mixed $^{13}$CO:$^{15}$N$_2$ (2:1) ice for which the photodesorption of $^{13}$CO at the $^{15}$N$_2$ N 1s $\rightarrow \pi^*$ transition (upper left panel of Figure \ref{fig:Yield_Mix}) and that of $^{15}$N$_2$ at the $^{13}$CO O 1s $\rightarrow \pi^*$ transition (lower right panel of Figure \ref{fig:Yield_Mix}) dominate the photodesorption spectrum of respectively $^{13}$CO and $^{15}$N$_2$. More globally, the fact that the photodesorption spectrum of $^{15}$N$_2$ and $^{13}$CO follows the X-ray absorption spectrum of, respectively, $^{13}$CO near the O K-edge and $^{15}$N$_2$ near the N K-edge confirms this indirect mechanism. Our experiment on the layered $^{13}$CO/$^{15}$N$_2$ ice also sheds light on this process: the ice is irradiated at 401 eV (N 1s $\rightarrow \pi^*$ transition of $^{15}$N$_2$) and $^{13}$CO desorption is observed.

This indirect desorption mechanism necessarily involves energy transport from the molecule that absorbs the incident photon to the molecule that desorbs from the ice surface. The length scale of such energy transport can be defined as the maximum distance from the ice surface (or the depth) at which an absorbed photon can induce desorption. This depth will be defined as $\Lambda_{des}$ in the following and it will be expressed in ML. $\Lambda_{des}$ depends on the indirect desorption mechanism and may vary with the photon energy and the ice composition. 

Our experiment on the layered $^{13}$CO/$^{15}$N$_2$ ice nicely enables to estimate $\Lambda_{des}$ for the indirect desorption process of $^{13}$CO induced by the photo-absorption of $^{15}$N$_2$ at 401 eV. In Figure \ref{fig:Layered_exp}, we can see that $^{15}$N$_2$ photo-absorption can induce $^{13}$CO desorption when the $^{15}$N$_2$ ice is covered by a number of ML of $^{13}$CO from 1 to roughly 30 - 40 ML, with a maximum efficiency around 10 ML. Above 40 ML, the photodesorption yield of $^{13}$CO from the layered ice is similar to the one induced by the photo-ionization of the core C 1s electron of $^{13}$CO at 530 eV in pure CO ice \citep{dupuy_co_2021} and the indirect desorption mechanism of $^{13}$CO induced by photo-absorption of $^{15}$N$_2$ is negligible. After fitting, the exponential decay of the $^{13}$CO photodesorption yield starting at $\sim$ 10 ML  is associated with a characteristic length of $\Lambda_c \sim$ 11 ML. The exponential reaches 95\% extinction at 3$\times \Lambda_c =$ 33 ML, which corresponds to a good estimate of $\Lambda_{des}$ and for which the indirect desorption of $^{13}$CO becomes negligible. In order to take into account the uncertainty associated with the number of ML of $^{13}$CO, which is approximately 10\%, we finally estimate that $\Lambda_{des}$ is between 30 - 40 ML in the case of our layered $^{13}$CO/$^{15}$N$_2$ ice and at 401 eV. This shows that desorption can be triggered from the bulk of the ice, beyond the first few ML. 

This conclusion can also be drawn for the case of mixed $^{13}$CO:$^{15}$N$_2$ ices. In that case, the photodesorption yield of $^{13}$CO at the N 1s $\rightarrow \pi^*$ transition of $^{15}$N$_2$ is exactly following the variations of the TEY from a 2:1 to a 20:1 ratio (see upper left panel of Figure \ref{fig:Yield_Mix}). This is not the case for the desorption of $^{15}$N$_2$ which is barely visible in the case of a 20:1 ratio (see lower left panel of Figure \ref{fig:Yield_Mix}). This could be explained as followed. In the case of the 2:1 ratio, $^{15}$N$_2$ molecules are sufficiently present near the ice surface and within its bulk to observe desorption of both $^{15}$N$_2$ and $^{13}$CO at the N 1s $\rightarrow \pi^*$ transition of $^{15}$N$_2$. Whereas, in the case of the 20:1 ratio, $^{15}$N$_2$ molecules are not sufficiently present near the ice surface to significantly desorb but they are sufficiently photo-absorbing from the surface to the bulk of the ice to produce a non-negligible TEY value and to induce the desorption of $^{13}$CO. In the case of mixed $^{13}$CO:$^{15}$N$_2$ ices, X-ray photodesorption can therefore also be triggered from the bulk of the ice.

\subsection{\label{sec:mechanisms} Indirect desorption mechanisms}

In this section, we discuss the possible mechanisms responsible for the indirect desorption of $^{13}$CO by excitation of $^{15}$N$_2$ observed in our layered ice experiment. Extrapolation to the case of the pure and mixed ices will be discussed afterwards. One possible mechanism which has been suggested for the X-ray photodesorption of neutrals from ices of simple (CO, H$_2$O) and more complex (CH$_3$OH) molecules \citep{dupuy_x-ray_2018, dupuy_desorption_2020,dupuy_co_2021, basalgete_complex_2021-1, basalgete_complex_2021} is X-ray induced electron stimulated desorption (XESD). XESD is a consequence of the relaxation of the core hole photo-excited or photo-ionized state which decays by emission of an Auger electron with a probability close to 1 for low Z elements \citep{Walters_1971, Krause_1979}. This Auger electron scatters inelastically in the ice, creating secondary events (e.g, ionizations and excitations) and a cascade of low energy (< 20 eV) secondary electrons that may lead in fine to desorption. 

On the other hand, mechanisms decorrelated from Auger scattering may also play a role. This include all the mechanisms related to the energy left to the photo-absorbing molecule after Auger decay. Quantitatively, most of the photon energy is given to the Auger electron in the form of kinetic energy ($\sim$ 380 eV for the N 1s $\rightarrow \pi^*$ transition of $^{15}$N$_2$; \cite{moddeman_determination_1971}) and a few tens of eV is left to the photo-absorbing molecule. This latter found itself in a singly or multiply valence-excited state, singly or doubly ionized depending on the energy of the incident photon. For instance, in the specific case of our layered ice experiment at 401 eV, $^{15}$N$_2$ is expected to be left, after Auger decay, in an excited ionized state ($^{15}$N$_2^+$)$^*$. The energy left to the molecule in the form of electronic excitation may be transported towards the ice surface via different mechanims to cause indirect desorption (i.e. desorption of other molecules). These mechanisms possibly include:
\begin{enumerate}[label={(\arabic*)}]
	\item 
	\begin{enumerate}[label={(\alph*)}, leftmargin=0.4cm]
	    \item indirect desorption due to migration of the electronic excitation through the ice (exciton migration). This mechanism concerns the transport of exciton in ices, a process that stems from the work of Kimmel and co-workers on electron-stimulated desorption (ESD) from D$_2$O ice \citep{Orlando_1997, Petrik_2003}. It has also been suggested to play a role in the ESD of C$_6$H$_6$ on top of water ice by migration of long-lived excitons through the ice \citep{marchione_efficient_2016, marchione_electron-induced_2017}	\item relaxation of the excited state to modes coupled with desorption similar to what is observed in VUV photodesorption experiments \citep{bertin_indirect_2013} for which it has been shown that excitation of $^{13}$CO (to the A$^1\Pi$ state) and $^{15}$N$_2$ (to the b$^1\Pi_u$ state) can induce the desorption of respectively $^{15}$N$_2$ and $^{13}$CO from layered ices. In that case, it was suggested an energy transfer between the excited and the desorbing molecule via excitation of collective vibrational modes (phonons). To simplify the following discussion, we will consider that this process occurs without exciton migration but one should note that it could take place after exciton migration (in that case, it is included in process (1)(a))
	\end{enumerate}    
	\item indirect desorption due to diffusion of energetic fragments if the ionized excited state is dissociative, similar to the kick-out mechanism suggested in the case of indirect desorption of simple molecules on top of water ice, induced by energetic H atoms that are produced from photodissociation of H$_2$O in the VUV range \citep{Dupuy_water_ind_2021}. In some extreme cases, the dissociation may occur before Auger decay
\end{enumerate}

The aim of our discussion is then to identify the main mechanism responsible for the indirect desorption observed in our layered $^{13}$CO/$^{15}$N$_2$ ice experiment. We will especially discriminate between the XESD process and the other processes cited previously that are decorrelated from the Auger scattering and discuss the processes with respect to the value derived for $\Lambda_{des}$ (30 - 40 ML). 

For a XESD process, the energy absorbed in the bulk of the ice by $^{15}$N$_2$ can be transported towards the ice surface by the scattering of the Auger electron. Considering the energy deposition profile of keV electrons in light materials (1 $\leq$ Z $\leq$ 18) from the work of \cite{valkealahti_energy_1989}, we can reasonably expect that the Auger electron and the secondary electrons can deposit their energy through a few tens of ML. In that sense, the value of $\Lambda_{des}$ previously estimated (30 - 40 ML) is consistent with a XESD process. The migration distance of excitons in water ice is potentially tens of ML \citep{Petrik_2003}. Migration of excitons in the case of weakly interacting Van der Waals solids such as in $^{15}$N$_2$ ice and migration between $^{15}$N$_2$ and $^{13}$CO layers has not been studied yet such that the process (1)(a) cannot be discussed regarding the value derived for $\Lambda_{des}$. However, as discussed in Section \ref{sec:comp_VUV}, we expect this process to depend on the exciton state hence on the incident photon energy. 
\begin{figure*}
\includegraphics[width=16cm]{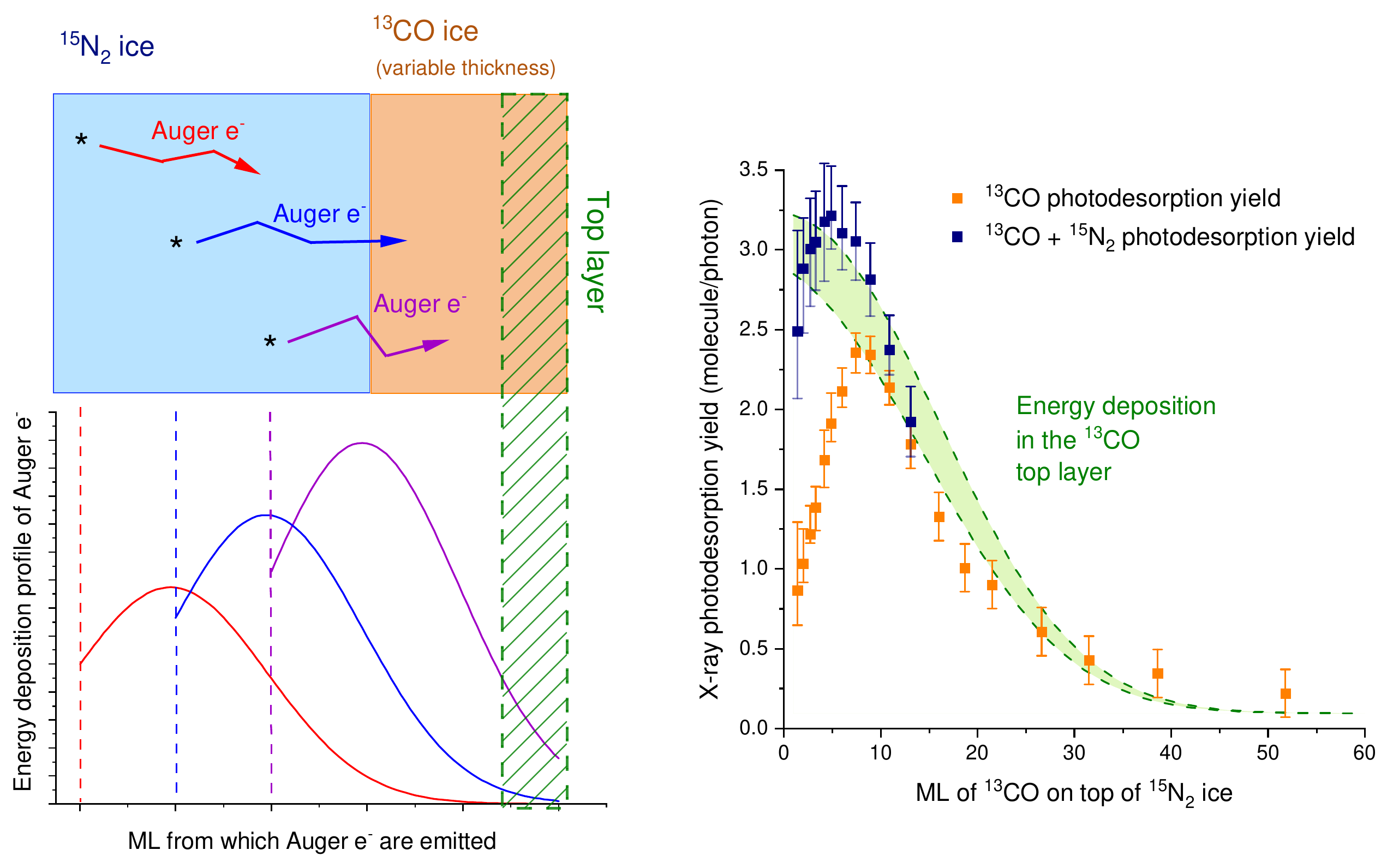}
\caption{\label{fig:eV_deposit} Left side: scheme of our simple model to estimate the energy deposited to the top $^{13}$CO layer of a layered $^{13}$CO/$^{15}$N$_2$ ice by scattering of Auger electrons emitted after photo-absorption by $^{15}$N$_2$ molecules. Right side: in green is displayed the estimation of this energy (in arbitrary units and where the zero of the right side distribution has been scaled with a photodesorption yield of 0.1 molecules photon$^{-1}$), compared with the X-ray photodesorption yields of $^{13}$CO (orange squares) from our layered $^{13}$CO/$^{15}$N$_2$ ice (from Figure \ref{fig:Layered_exp}). The blue squares show the sum of the photodesorption yields of $^{13}$CO plus $^{15}$N$_2$ from our layered $^{13}$CO/$^{15}$N$_2$ ice. The upper and lower green curve were computed by taking a photoabsorption cross section at 401 eV of 40 and 20 Mbarn respectively.}
\end{figure*}

Process (1)(b) has been characterized by \cite{bertin_indirect_2013} for the indirect desorption of $^{13}$CO and $^{15}$N$_2$ by excitation of respectively $^{15}$N$_2$ and $^{13}$CO in layered ice experiments. In that case, it has been shown that the energy can be efficiently transferred over $\sim$ 5 ML to the desorbing molecules at the surface (this value was similar for excitation of $^{13}$CO to the A$^1\Pi$ state and for excitation of $^{15}$N$_2$ to the b$^1\Pi_u$ state). Therefore, we expect process (1)(b) to be a near-surface process. Process (2) involves the diffusion of ionic or neutral fragments (in that case N or N$^+$) towards the ice surface and collision with $^{13}$CO molecules followed by desorption. After possible dissociation of ($^{15}$N$_2^+$)$^*$, we expect the fragments to have less than 10 eV of kinetic energy. Although no studies are available on the typical diffusion length of such fragments in the bulk of molecular ices, we do not expect it to be comparable with the value of $\Lambda_{des}$ derived. Instead, we expect process (2) to be a near-surface process involving just the first few ML of the ice. This is also supported by the very low level of desorbing $^{15}$N fragments from pure $^{15}$N$_2$ ice (see section \ref{sec:Yields_pure}). 

Processes (1)(b) and (2) are then expected to possibly occur in the low coverage regime of our layered ice experiment, for a few ML of $^{13}$CO on top of the $^{15}$N$_2$ ice. Regarding the photodesorption yields of $^{13}$CO in Figure \ref{fig:Layered_exp}, these near-surface processes seem to be negligible compared to the processes involving deeper layers of the $^{15}$N$_2$ ice, such as XESD. In fact, the yields of $^{13}$CO at low coverage (between 1 and 10 ML) and higher coverage are similar. In the low coverage regime, both near-surface processes and bulk processes can participate to indirect desorption whereas for higher coverage, only bulk processes participate. The fact that the yields are similar in both regimes excludes a significant contribution of the near-surface processes compared to the bulk ones. Therefore, processes (1)(b) and (2) can be neglected. 

In the following sections, we focus on characterizing the XESD process for our layered ice experiment and we conclude on the dominant mechanism expected for indirect desorption of neutral $^{15}$N$_2$ and $^{13}$CO from the ices studied in this work. 

\subsection{\label{sec:model}Modeling of the energy deposited in the ice by Auger electrons}

XESD is mediated by the scattering of Auger electrons. These electrons deposit their energy through the ice via ionization and excitation of the molecules in their path. For a XESD process, the photodesorption yields are expected to be correlated to the amount of energy deposited near the ice surface by the scattering of Auger electrons. This amount of energy can be estimated based on the work of \cite{valkealahti_energy_1989}. They showed, by Monte Carlo simulations, that the energy deposition profile of a keV electron in a light material (1 $\leq$ Z $\leq$ 18) as a function of depth can be approximated by a Gaussian distribution (in 1D). The parameters of the distribution depend on the energy of the electron and on the stopping power of the material at this energy. These parameters have been fitted for a N$_2$ ice. We can therefore estimate the amount of energy deposited near the ice surface by the scattering of Auger electrons in the case of our layered $^{13}$CO/$^{15}$N$_2$ ice experiment using the following methodology and assumptions (this is also sketched in the left side of Figure \ref{fig:eV_deposit}):
\begin{itemize}
 \item we consider a layered $^{13}$CO/$^{15}$N$_2$ ice: 50 ML of $^{15}$N$_2$ on top of which an increasing number of ML of $^{13}$CO is deposited. We neglect the photo-absorption of the $^{13}$CO layers at 401 eV
 \item we assume that the initial kinetic energy of the Auger electron emitted by relaxation of the N (1s)$^{-1}\pi^*$ state of $^{15}$N$_2$ is $\sim$ 380 eV, based on Auger spectra of gas phase N$_2$ \citep{moddeman_determination_1971} 
 \item we only consider the emitted Auger electrons that statistically contribute the most to the energy deposited near the ice surface, i.e. the Auger electrons that are emitted from the bulk towards the surface of the ice
 \item the energy deposition profile of the Auger electrons is approximated by a Gaussian distribution with parameters from \cite{valkealahti_energy_1989}, in 1D. This also takes into account the energy loss of secondary electrons and it is computed as the sum of the average kinetic energy of all electrons in a depth element. According to the Born-Bethe approximation and Bragg's rule, the stopping power of $^{13}$CO and $^{15}$N$_2$ at 380 eV are similar such that we reasonably assume that the energy deposition profile does not significantly differs in $^{15}$N$_2$ or $^{13}$CO layers. The shape of the energy deposition profiles are plotted in arbitrary units in the left panel of Figure \ref{fig:eV_deposit} 
 \item we consider that each ML of $^{15}$N$_2$ is emitting a number of Auger electrons that are equal to the number of photons absorbed in that ML. It results that the relative intensity of the energy deposition distributions from two different ML is driven by a Beer-Lambert law, with an absorption cross section at 401 eV taken between 20 and 40 Mbarn \citep{kato_absolute_2007}
\end{itemize}

The energy deposited in the top layer of the $^{13}$CO ice is then computed by summing the contributions from each ML of $^{15}$N$_2$. This operation is repeated for an increasing number of ML of $^{13}$CO deposited on top of the $^{15}$N$_2$ ice. The results are presented in the right panel of Figure \ref{fig:eV_deposit}: the energy deposited on the top layer of the $^{13}$CO ice is compared to the experimental photodesorption yields of $^{13}$CO from our layered $^{13}$CO/$^{15}$N$_2$ ice (taken from Figure \ref{fig:Layered_exp}). A comparison with the sum of the photodesorption yields of $^{13}$CO plus $^{15}$N$_2$, which have similar photodesorption yields at 401 eV when mixed in similar stoichiometries (according to our results from mixed ices), is also done for the following discussion. At low coverage (number of ML of $^{13}$CO $\lesssim$ 10 ML), the ice surface is not homogeneously covered by $^{13}$CO molecules and the photodesorption yield of $^{13}$CO increases linearly with the number of $^{13}$CO molecules available for desorption. Therefore, in this regime, the photodesorption yield of $^{13}$CO is not correlated to the energy deposition of Auger electrons since it actually depends on the ice surface homogeneity. This lack of homogeneity is supported by the fact that $^{15}$N$_2$ still desorbs in this regime. Furthermore, in this low coverage regime, we observe a good agreement between the energy deposition profile and the sum of the photodesorption yields of $^{13}$CO plus $^{15}$N$_2$ (see blue squares in Figure \ref{fig:eV_deposit}). This is consistent with the ice surface being composed of a mix of $^{13}$CO and $^{15}$N$_2$ molecules in the low coverage regime and for which photodesorption of neutral molecules is correlated to the energy deposited at the top layer and dominated by $^{13}$CO plus $^{15}$N$_2$ desorption (no significant desorption is found for other possible neutral molecules). For medium coverage (number of ML of $^{13}$CO between 10 and 40 ML), the exponential decay of the $^{13}$CO desorption yield is in good agreement with the modeled deposited energy. For high coverage (number of ML of $^{13}$CO $>$ 40 ML), the photodesorption is triggered by the photo-ionization of $^{13}$CO (see Section \ref{sec:Layered}).

These results strongly suggest that XESD is the dominant mechanism responsible for the indirect desorption of $^{13}$CO at 401 eV from our layered $^{13}$CO/$^{15}$N$_2$ ice. It also confirms our estimated value for $\Lambda_{des}$. In that case, excitation of $^{13}$CO to non-dissociative states by low energy secondary electrons (produced by the Auger scattering) may be the exact mechanism explaining $^{13}$CO desorption. For comparison, ESD of CO from pure CO ice has an energy threshold around 6 eV \citep{RAKHOVSKAIA_1995}, which is accessible to the secondary electrons, as suggested in \cite{dupuy_co_2021}. This energy threshold is around 7 eV for the ESD of N$_2$ from pure N$_2$ ice. 

Auger emission occurs with a probability close to one for each core hole state of $^{15}$N$_2$ or $^{13}$CO above and below their respective IP. The energy deposition profile due to the Auger scattering is also reasonably expected to be similar in the different ices tested in this work (pure $^{15}$N$_2$ ice, mixed $^{13}$CO:$^{15}$N$_2$ ices and layered $^{13}$CO/$^{15}$N$_2$ ice), based on the fact that $^{13}$CO and $^{15}$N$_2$ have similar stopping power in the energy range considered. It is therefore straightforward to consider that XESD should also play a role for the desorption of $^{15}$N$_2$ from pure $^{15}$N$_2$ ice and for the desorption of $^{13}$CO and $^{15}$N$_2$ from mixed $^{13}$CO:$^{15}$N$_2$ ices, at each photon energy. The simple model previously implemented nicely enables to estimate $\Lambda_{des}$ for a XESD process at different photon energy as long as the photo-absorption cross section and the initial kinetic energy of the Auger electron emitted after core hole decay are known. This is used in the next section.

{\renewcommand{\arraystretch}{2}
\begin{table*}[t!]
\begin{threeparttable}
\caption{\label{tab:yield_eV}Photodesorption yields expressed in molecule desorbed by eV deposited, of $^{13}$CO and $^{15}$N$_2$ from different ices and at different energies, computed via equation (\ref{eq:gamma_ev}) and by using the experimental yields derived in this work. The lower and higher values of the yields are derived by taking the higher and lower value of $\Lambda_{des}$ respectively.}
\begin{ruledtabular}
\begin{tabular}{lcccccc}
 &&&\multicolumn{4}{c}{Photodesorption yield ($\times 10^{-2}$ molecules (eV deposited)$^{-1}$) } \\ 
 
 &&&\multicolumn{2}{c}{$^{15}$N$_2$ from}&\multicolumn{2}{c}{$^{13}$CO from}\\\cline{4-5} \cline{6-7}
 
 Photon Energy & $\sigma$ (Mbarn)$^{(a)}$ & $\Lambda_{des}$ (ML) & Pure $^{15}$N$_2$ & $^{13}$CO:$^{15}$N$_2$ (2:1) & Pure $^{13}$CO & $^{13}$CO:$^{15}$N$_2$ (2:1)\\
 \hline

 401 eV ($^{15}$N$_2$; N 1s $\rightarrow \pi^*$) & 26 & 30 - 40 & 1.9 - 2.3 & 3.2 - 4.1 & & 1.9 - 2.5 \\
 
 425 eV ($^{15}$N$_2$; N 1s ionization) & 1.8 & 30 - 40 & 0.6 - 0.8 & (*) & & 2.1 - 2.8 \\
 
 534.2 eV ($^{13}$CO; O 1s $\rightarrow \pi^*$) & 3.0 & 40 - 50 & & 2.8 - 3.5 & & 2.1 - 2.6 \\

 570 eV ($^{13}$CO; O 1s ionization) & 0.8 & 40 - 50 & & (*) & 1.5$^{(b)}$ & 2.4 - 3.0 \\\hline
 
 8.3 eV ($^{13}$CO; A$^1 \Pi$ - X$^1 \Sigma^+ \ (2,0)$)$^{(c)}$ & 15 & 5 & & 18 & 8.3 & 21 \\

 \end{tabular}
 \end{ruledtabular}
 \begin{tablenotes}[flushleft]
 \item[(*)]Background issues on the mass channel 30 around 425 eV and 570 eV during the X-ray irradiation of the $^{13}$CO:$^{15}$N$_2$ -2:1 ice prevent us from using the experimental yields of $^{15}$N$_2$ at these energies
 \item[(a)]The photo-absorption cross section are taken from gas phase measurements \citep{barrus_k_1979, kato_absolute_2007} and we took into account the spectral resolution of our photon source.
 \item[(b)]Taken from \cite{dupuy_co_2021} where it was assumed $\Lambda_{des} = $ 30 ML
 \item[(c)]From \cite{bertin_indirect_2013}
\end{tablenotes}
\end{threeparttable}
\end{table*}
}

\subsection{\label{sec:comp_VUV}Conversion of the absorbed energy to desorption}
 
 In this section, we correct our X-ray photodesorption yields by taking into account only the part of the absorbed energy that can participate to the desorption process. This results in expressing the yields in molecules (eV deposited)$^{-1}$ using equation (\ref{eq:gamma_ev}). We consider that the number of the top monolayers involved in the desorption process ($\Lambda_{des}$) should be at least equal to the one corresponding to a XESD process. The energy absorbed in the layers beyond $\Lambda_{des}$ is "lost" in the sense that it cannot be converted to desorption. This results in taking $\Lambda_{des} =$ 30 - 40 ML near the N K-edge, where the photo-absorption is dominated by $^{15}$N$_2$ absorption. Near the O K-edge, the photo-absorption is dominated by $^{13}$CO absorption and the Auger electron is expected to have an initial kinetic energy between 500 and 520 eV \citep{moddeman_determination_1971}. Based on our simple modeling in Section \ref{sec:model}, this results in a slightly higher $\Lambda_{des}$ (40 - 50 ML). The results are presented in Table \ref{tab:yield_eV} (first four rows) for the desorption of $^{13}$CO and $^{15}$N$_2$ from their respective pure ice and from mixed $^{13}$CO:$^{15}$N$_2$ (2:1) ices at different photon energy. 

The yields displayed in Table \ref{tab:yield_eV} represent a quantitative estimate of how efficiently the absorbed energy (only the part that can participate to the desorption process) is converted to the desorption channel. This takes into account all the desorption processes that may occur due to photo-absorption in the first $\Lambda_{des}$ ML of the ice. The estimated yields are very similar for the desorption of $^{13}$CO and $^{15}$N$_2$ from pure or mixed ices and at each energy considered. They range between $\sim$ 0.02 and $\sim$ 0.04 molecules (eV deposited)$^{-1}$. There is however a small discrepancy with the photodesorption yield of $^{15}$N$_2$ from pure $^{15}$N$_2$ ice at 425 eV, which is lower than the previous values. This is assumed to be due to a higher fluence received by the ice before this specific measurement compared to the other ones. 

These results indicate that there is presumably a unique dominant process explaining the desorption of $^{13}$CO and $^{15}$N$_2$ from the ices tested and whose efficiency does not strongly depend on the photon energy hence neither on the photo-absorbing molecule nor on its state after Auger decay (singly or doubly ionized excited state). This is fully consistent with a XESD process for which desorption depends only on the Auger scattering, the latter being expected to not significantly differ between pure and mixed ices of $^{13}$CO and $^{15}$N$_2$. The similarity of the yields between $^{13}$CO and $^{15}$N$_2$ is also consistent with their similar binding energies in pure or mixed ice \citep{oberg_competition_2005, bisschop_desorption_2006}. A possible significant desorption due to exciton migration from the photo-absorbing molecule (process (1)(a) defined in Section \ref{sec:mechanisms}) can also be excluded by the fact that the efficiency of this process is expected to depend on the exciton state after Auger decay and therefore on the photon energy, which is not what we observe in Table \ref{tab:yield_eV}.  

Finally, in Table \ref{tab:yield_eV}, we also computed the VUV photodesorption yields in molecules (eV deposited)$^{-1}$ at 8.3 eV, which corresponds to the A$^1 \Pi$ - X$^1 \Sigma^+ \ (2,0)$ electronic transition of $^{13}$CO, from the data published in \cite{bertin_indirect_2013}. In this study, $\Lambda_{des}$ was estimated to be $\sim$ 5 ML for the indirect desorption mechanism from layered $^{15}$N$_2$/$^{13}$CO experiments. The yields in molecules (eV deposited)$^{-1}$ are found higher in the VUV range (at 8.3 eV) than in the X-ray range. This means that the energy absorbed by a valence excitation of $^{13}$CO (to the A$^{1}\Pi$ state), is more efficiently converted to desorption, by a process similar to the one described in Section \ref{sec:mechanisms} (process (1)(b)), than the energy absorbed by a core hole excitation or ionization (of $^{13}$CO or $^{15}$N$_2$) and for which a XESD process is expected to be dominant according to our results. This is true for both the VUV photodesorption of $^{13}$CO (direct and/or indirect process) and $^{15}$N$_2$ (indirect process) from pure $^{13}$CO ice and mixed $^{13}$CO:$^{15}$N$_2$ ice.

\section{Conclusion}

X-ray photodesorption of $^{15}$N$_2$ from pure $^{15}$N$_2$ ice and of $^{15}$N$_2$ and $^{13}$CO from mixed $^{13}$CO:$^{15}$N$_2$ ices is found to follow the X-ray absorption profile of the ices in the N and O K-edge energy range (near 400 and 500 eV respectively). Photodesorption experiments on mixed $^{13}$CO:$^{15}$N$_2$ ices revealed an indirect desorption mechanism triggered by the absorption of $^{15}$N$_2$ near the N K-edge and $^{13}$CO near the O K-edge leading both to $^{13}$CO and $^{15}$N$_2$ desorption. This indirect mechanism is also highlighted by the desorption of $^{13}$CO from a X-ray irradiated layered $^{13}$CO/$^{15}$N$_2$ ice at 401 eV. This latter experiment enables to derive the relevant depth of desorption $\Lambda_{des}$ which is a quantitative measure of depth involved in the indirect desorption process. It was estimated to be 30 - 40 ML near the N K-edge. The photodesorption yields of $^{13}$CO from the layered $^{13}$CO/$^{15}$N$_2$ ice as a function of $^{13}$CO coverage are found to be well-correlated to the amount of energy deposited at the top layer by the scattering of Auger electrons after X-ray absorption, the latter being estimated by a simple model based on the work of \cite{valkealahti_energy_1989}. The desorption yields of $^{13}$CO and $^{15}$N$_2$ corrected from the absorbed energy that can participate to desorption do not significantly depend on the photon energy (hence on the photo-absorbing molecule, $^{13}$CO or $^{15}$N$_2$) and on the nature of the ice (pure or mixed). These results indicate that XESD, mediated by the scattering of Auger electrons, is the dominant mechanism explaining the X-ray photodesorption of $^{15}$N$_2$ and $^{13}$CO from the ices tested in this work. 

\section*{Acknowledgements}
This work was done with financial support from (i) the Region Ile-de-France DIM-ACAV+ program, (ii) the Sorbonne Université “Emergence” program, (iii) the ANR PIXyES project, grant ANR-20-CE30-0018 of the French “Agence Nationale de la Recherche,” and (iv) the Programme National “Physique et Chimie du Milieu Interstellaire” (PCMI) of CNRS/INSU with INC/INP co-funded by CEA and CNES. We would like to acknowledge SOLEIL for provision of synchrotron radiation facilities under Project Nos. 20210142, and we thank N. Jaouen, H. Popescu and R. Gaudemer for their help on the SEXTANTS beam
line. 

\section*{Notes and References}
\bibliography{MyBibli}

\begin{thebibliography}{50}%
\makeatletter
\providecommand \@ifxundefined [1]{%
 \@ifx{#1\undefined}
}%
\providecommand \@ifnum [1]{%
 \ifnum #1\expandafter \@firstoftwo
 \else \expandafter \@secondoftwo
 \fi
}%
\providecommand \@ifx [1]{%
 \ifx #1\expandafter \@firstoftwo
 \else \expandafter \@secondoftwo
 \fi
}%
\providecommand \natexlab [1]{#1}%
\providecommand \enquote  [1]{``#1''}%
\providecommand \bibnamefont  [1]{#1}%
\providecommand \bibfnamefont [1]{#1}%
\providecommand \citenamefont [1]{#1}%
\providecommand \href@noop [0]{\@secondoftwo}%
\providecommand \href [0]{\begingroup \@sanitize@url \@href}%
\providecommand \@href[1]{\@@startlink{#1}\@@href}%
\providecommand \@@href[1]{\endgroup#1\@@endlink}%
\providecommand \@sanitize@url [0]{\catcode `\\12\catcode `\$12\catcode
  `\&12\catcode `\#12\catcode `\^12\catcode `\_12\catcode `\%12\relax}%
\providecommand \@@startlink[1]{}%
\providecommand \@@endlink[0]{}%
\providecommand \url  [0]{\begingroup\@sanitize@url \@url }%
\providecommand \@url [1]{\endgroup\@href {#1}{\urlprefix }}%
\providecommand \urlprefix  [0]{URL }%
\providecommand \Eprint [0]{\href }%
\providecommand \doibase [0]{http://dx.doi.org/}%
\providecommand \selectlanguage [0]{\@gobble}%
\providecommand \bibinfo  [0]{\@secondoftwo}%
\providecommand \bibfield  [0]{\@secondoftwo}%
\providecommand \translation [1]{[#1]}%
\providecommand \BibitemOpen [0]{}%
\providecommand \bibitemStop [0]{}%
\providecommand \bibitemNoStop [0]{.\EOS\space}%
\providecommand \EOS [0]{\spacefactor3000\relax}%
\providecommand \BibitemShut  [1]{\csname bibitem#1\endcsname}%
\let\auto@bib@innerbib\@empty
\bibitem [{\citenamefont {Barrus}\ \emph {et~al.}(1979)\citenamefont {Barrus},
  \citenamefont {Blake}, \citenamefont {Burek}, \citenamefont {Chambers},\ and\
  \citenamefont {Pregenzer}}]{barrus_k_1979}%
  \BibitemOpen
  \bibfield  {author} {\bibinfo {author} {\bibnamefont {Barrus}, \bibfnamefont
  {D.~M.}}, \bibinfo {author} {\bibnamefont {Blake}, \bibfnamefont {R.~L.}},
  \bibinfo {author} {\bibnamefont {Burek}, \bibfnamefont {A.~J.}}, \bibinfo
  {author} {\bibnamefont {Chambers}, \bibfnamefont {K.~C.}}, \ and\ \bibinfo
  {author} {\bibnamefont {Pregenzer}, \bibfnamefont {A.~L.}},\ }\bibfield
  {title} {\enquote {\bibinfo {title} {K -shell photoabsorption coefficients of
  {O} 2 , {C} {O} 2 , {CO}, and {N} 2 {O}},}\ }\href {\doibase
  10.1103/PhysRevA.20.1045} {\bibfield  {journal} {\bibinfo  {journal}
  {Physical Review A}\ }\textbf {\bibinfo {volume} {20}},\ \bibinfo {pages}
  {1045--1061} (\bibinfo {year} {1979})}\BibitemShut {NoStop}%
\bibitem [{\citenamefont {Basalgète}\ \emph
  {et~al.}(2021{\natexlab{a}})\citenamefont {Basalgète}, \citenamefont
  {Dupuy}, \citenamefont {Féraud}, \citenamefont {Romanzin}, \citenamefont
  {Philippe}, \citenamefont {Michaut}, \citenamefont {Michoud}, \citenamefont
  {Amiaud}, \citenamefont {Lafosse}, \citenamefont {Fillion},\ and\
  \citenamefont {Bertin}}]{basalgete_complex_2021-1}%
  \BibitemOpen
  \bibfield  {author} {\bibinfo {author} {\bibnamefont {Basalgète},
  \bibfnamefont {R.}}, \bibinfo {author} {\bibnamefont {Dupuy}, \bibfnamefont
  {R.}}, \bibinfo {author} {\bibnamefont {Féraud}, \bibfnamefont {G.}},
  \bibinfo {author} {\bibnamefont {Romanzin}, \bibfnamefont {C.}}, \bibinfo
  {author} {\bibnamefont {Philippe}, \bibfnamefont {L.}}, \bibinfo {author}
  {\bibnamefont {Michaut}, \bibfnamefont {X.}}, \bibinfo {author} {\bibnamefont
  {Michoud}, \bibfnamefont {J.}}, \bibinfo {author} {\bibnamefont {Amiaud},
  \bibfnamefont {L.}}, \bibinfo {author} {\bibnamefont {Lafosse}, \bibfnamefont
  {A.}}, \bibinfo {author} {\bibnamefont {Fillion}, \bibfnamefont {J.-H.}}, \
  and\ \bibinfo {author} {\bibnamefont {Bertin}, \bibfnamefont {M.}},\
  }\bibfield  {title} {\enquote {\bibinfo {title} {Complex organic molecules in
  protoplanetary disks: {X}-ray photodesorption from methanol-containing ices:
  {I}. {Pure} methanol ices},}\ }\href {\doibase 10.1051/0004-6361/202039676}
  {\bibfield  {journal} {\bibinfo  {journal} {Astronomy \& Astrophysics}\
  }\textbf {\bibinfo {volume} {647}},\ \bibinfo {pages} {A35} (\bibinfo {year}
  {2021}{\natexlab{a}})}\BibitemShut {NoStop}%
\bibitem [{\citenamefont {Basalgète}\ \emph
  {et~al.}(2021{\natexlab{b}})\citenamefont {Basalgète}, \citenamefont
  {Dupuy}, \citenamefont {Féraud}, \citenamefont {Romanzin}, \citenamefont
  {Philippe}, \citenamefont {Michaut}, \citenamefont {Michoud}, \citenamefont
  {Amiaud}, \citenamefont {Lafosse}, \citenamefont {Fillion},\ and\
  \citenamefont {Bertin}}]{basalgete_complex_2021}%
  \BibitemOpen
  \bibfield  {author} {\bibinfo {author} {\bibnamefont {Basalgète},
  \bibfnamefont {R.}}, \bibinfo {author} {\bibnamefont {Dupuy}, \bibfnamefont
  {R.}}, \bibinfo {author} {\bibnamefont {Féraud}, \bibfnamefont {G.}},
  \bibinfo {author} {\bibnamefont {Romanzin}, \bibfnamefont {C.}}, \bibinfo
  {author} {\bibnamefont {Philippe}, \bibfnamefont {L.}}, \bibinfo {author}
  {\bibnamefont {Michaut}, \bibfnamefont {X.}}, \bibinfo {author} {\bibnamefont
  {Michoud}, \bibfnamefont {J.}}, \bibinfo {author} {\bibnamefont {Amiaud},
  \bibfnamefont {L.}}, \bibinfo {author} {\bibnamefont {Lafosse}, \bibfnamefont
  {A.}}, \bibinfo {author} {\bibnamefont {Fillion}, \bibfnamefont {J.-H.}}, \
  and\ \bibinfo {author} {\bibnamefont {Bertin}, \bibfnamefont {M.}},\
  }\bibfield  {title} {\enquote {\bibinfo {title} {Complex organic molecules in
  protoplanetary disks: {X}-ray photodesorption from methanol-containing ices:
  {II}. {Mixed} methanol-{CO} and methanol-{H} $_{\textrm{2}}$ {O} ices},}\
  }\href {\doibase 10.1051/0004-6361/202040117} {\bibfield  {journal} {\bibinfo
   {journal} {Astronomy \& Astrophysics}\ }\textbf {\bibinfo {volume} {647}},\
  \bibinfo {pages} {A36} (\bibinfo {year} {2021}{\natexlab{b}})}\BibitemShut
  {NoStop}%
\bibitem [{\citenamefont {Bennett}, \citenamefont {Pirim},\ and\ \citenamefont
  {Orlando}(2013)}]{Bennet_2013}%
  \BibitemOpen
  \bibfield  {author} {\bibinfo {author} {\bibnamefont {Bennett}, \bibfnamefont
  {C.~J.}}, \bibinfo {author} {\bibnamefont {Pirim}, \bibfnamefont {C.}}, \
  and\ \bibinfo {author} {\bibnamefont {Orlando}, \bibfnamefont {T.~M.}},\
  }\bibfield  {title} {\enquote {\bibinfo {title} {Space-weathering of solar
  system bodies: A laboratory perspective},}\ }\href {\doibase
  10.1021/cr400153k} {\bibfield  {journal} {\bibinfo  {journal} {Chemical
  Reviews}\ }\textbf {\bibinfo {volume} {113}},\ \bibinfo {pages} {9086--9150}
  (\bibinfo {year} {2013})}\BibitemShut {NoStop}%
\bibitem [{\citenamefont {Bertin}\ \emph {et~al.}(2013)\citenamefont {Bertin},
  \citenamefont {Fayolle}, \citenamefont {Romanzin}, \citenamefont {Poderoso},
  \citenamefont {Michaut}, \citenamefont {Philippe}, \citenamefont {Jeseck},
  \citenamefont {Öberg}, \citenamefont {Linnartz},\ and\ \citenamefont
  {Fillion}}]{bertin_indirect_2013}%
  \BibitemOpen
  \bibfield  {author} {\bibinfo {author} {\bibnamefont {Bertin}, \bibfnamefont
  {M.}}, \bibinfo {author} {\bibnamefont {Fayolle}, \bibfnamefont {E.~C.}},
  \bibinfo {author} {\bibnamefont {Romanzin}, \bibfnamefont {C.}}, \bibinfo
  {author} {\bibnamefont {Poderoso}, \bibfnamefont {H.~A.~M.}}, \bibinfo
  {author} {\bibnamefont {Michaut}, \bibfnamefont {X.}}, \bibinfo {author}
  {\bibnamefont {Philippe}, \bibfnamefont {L.}}, \bibinfo {author}
  {\bibnamefont {Jeseck}, \bibfnamefont {P.}}, \bibinfo {author} {\bibnamefont
  {Öberg}, \bibfnamefont {K.~I.}}, \bibinfo {author} {\bibnamefont {Linnartz},
  \bibfnamefont {H.}}, \ and\ \bibinfo {author} {\bibnamefont {Fillion},
  \bibfnamefont {J.-H.}},\ }\bibfield  {title} {\enquote {\bibinfo {title}
  {{INDIRECT} {ULTRAVIOLET} {PHOTODESORPTION} {FROM} {CO}:{N} $_{\textrm{2}}$
  {BINARY} {ICES} — {AN} {EFFICIENT} {GRAIN}-{GAS} {PROCESS}},}\ }\href
  {\doibase 10.1088/0004-637X/779/2/120} {\bibfield  {journal} {\bibinfo
  {journal} {The Astrophysical Journal}\ }\textbf {\bibinfo {volume} {779}},\
  \bibinfo {pages} {120} (\bibinfo {year} {2013})}\BibitemShut {NoStop}%
\bibitem [{\citenamefont {Bertin}\ \emph {et~al.}(2012)\citenamefont {Bertin},
  \citenamefont {Fayolle}, \citenamefont {Romanzin}, \citenamefont {Öberg},
  \citenamefont {Michaut}, \citenamefont {Moudens}, \citenamefont {Philippe},
  \citenamefont {Jeseck}, \citenamefont {Linnartz},\ and\ \citenamefont
  {Fillion}}]{bertin_uv_2012}%
  \BibitemOpen
  \bibfield  {author} {\bibinfo {author} {\bibnamefont {Bertin}, \bibfnamefont
  {M.}}, \bibinfo {author} {\bibnamefont {Fayolle}, \bibfnamefont {E.~C.}},
  \bibinfo {author} {\bibnamefont {Romanzin}, \bibfnamefont {C.}}, \bibinfo
  {author} {\bibnamefont {Öberg}, \bibfnamefont {K.~I.}}, \bibinfo {author}
  {\bibnamefont {Michaut}, \bibfnamefont {X.}}, \bibinfo {author} {\bibnamefont
  {Moudens}, \bibfnamefont {A.}}, \bibinfo {author} {\bibnamefont {Philippe},
  \bibfnamefont {L.}}, \bibinfo {author} {\bibnamefont {Jeseck}, \bibfnamefont
  {P.}}, \bibinfo {author} {\bibnamefont {Linnartz}, \bibfnamefont {H.}}, \
  and\ \bibinfo {author} {\bibnamefont {Fillion}, \bibfnamefont {J.-H.}},\
  }\bibfield  {title} {\enquote {\bibinfo {title} {{UV} photodesorption of
  interstellar {CO} ice analogues: from subsurface excitation to surface
  desorption},}\ }\href {\doibase 10.1039/c2cp41177f} {\bibfield  {journal}
  {\bibinfo  {journal} {Physical Chemistry Chemical Physics}\ }\textbf
  {\bibinfo {volume} {14}},\ \bibinfo {pages} {9929} (\bibinfo {year}
  {2012})}\BibitemShut {NoStop}%
\bibitem [{\citenamefont {Bisschop}\ \emph {et~al.}(2006)\citenamefont
  {Bisschop}, \citenamefont {Fraser}, \citenamefont {Öberg}, \citenamefont
  {van Dishoeck},\ and\ \citenamefont {Schlemmer}}]{bisschop_desorption_2006}%
  \BibitemOpen
  \bibfield  {author} {\bibinfo {author} {\bibnamefont {Bisschop},
  \bibfnamefont {S.~E.}}, \bibinfo {author} {\bibnamefont {Fraser},
  \bibfnamefont {H.~J.}}, \bibinfo {author} {\bibnamefont {Öberg},
  \bibfnamefont {K.~I.}}, \bibinfo {author} {\bibnamefont {van Dishoeck},
  \bibfnamefont {E.~F.}}, \ and\ \bibinfo {author} {\bibnamefont {Schlemmer},
  \bibfnamefont {S.}},\ }\bibfield  {title} {\enquote {\bibinfo {title}
  {Desorption rates and sticking coefficients for {CO} and {N} $_{\textrm{2}}$
  interstellar ices},}\ }\href {\doibase 10.1051/0004-6361:20054051} {\bibfield
   {journal} {\bibinfo  {journal} {Astronomy \& Astrophysics}\ }\textbf
  {\bibinfo {volume} {449}},\ \bibinfo {pages} {1297--1309} (\bibinfo {year}
  {2006})}\BibitemShut {NoStop}%
\bibitem [{\citenamefont {Boogert}, \citenamefont {Gerakines},\ and\
  \citenamefont {Whittet}(2015)}]{boogert_observations_2015}%
  \BibitemOpen
  \bibfield  {author} {\bibinfo {author} {\bibnamefont {Boogert}, \bibfnamefont
  {A.~A.}}, \bibinfo {author} {\bibnamefont {Gerakines}, \bibfnamefont
  {P.~A.}}, \ and\ \bibinfo {author} {\bibnamefont {Whittet}, \bibfnamefont
  {D.~C.}},\ }\bibfield  {title} {\enquote {\bibinfo {title} {Observations of
  the {Icy} {Universe}},}\ }\href {\doibase
  10.1146/annurev-astro-082214-122348} {\bibfield  {journal} {\bibinfo
  {journal} {Annual Review of Astronomy and Astrophysics}\ }\textbf {\bibinfo
  {volume} {53}},\ \bibinfo {pages} {541--581} (\bibinfo {year}
  {2015})}\BibitemShut {NoStop}%
\bibitem [{\citenamefont {Chen}, \citenamefont {Ma},\ and\ \citenamefont
  {Sette}(1989)}]{chen_k_1989}%
  \BibitemOpen
  \bibfield  {author} {\bibinfo {author} {\bibnamefont {Chen}, \bibfnamefont
  {C.~T.}}, \bibinfo {author} {\bibnamefont {Ma}, \bibfnamefont {Y.}}, \ and\
  \bibinfo {author} {\bibnamefont {Sette}, \bibfnamefont {F.}},\ }\bibfield
  {title} {\enquote {\bibinfo {title} {\textit{{K}} -shell photoabsorption of
  the {N} 2 molecule},}\ }\href {\doibase 10.1103/PhysRevA.40.6737} {\bibfield
  {journal} {\bibinfo  {journal} {Physical Review A}\ }\textbf {\bibinfo
  {volume} {40}},\ \bibinfo {pages} {6737--6740} (\bibinfo {year}
  {1989})}\BibitemShut {NoStop}%
\bibitem [{\citenamefont {Ciaravella}\ \emph {et~al.}(2020)\citenamefont
  {Ciaravella}, \citenamefont {Muñoz~Caro}, \citenamefont {Jiménez-Escobar},
  \citenamefont {Cecchi-Pestellini}, \citenamefont {Hsiao}, \citenamefont
  {Huang},\ and\ \citenamefont {Chen}}]{ciaravella_x-ray_2020}%
  \BibitemOpen
  \bibfield  {author} {\bibinfo {author} {\bibnamefont {Ciaravella},
  \bibfnamefont {A.}}, \bibinfo {author} {\bibnamefont {Muñoz~Caro},
  \bibfnamefont {G.~M.}}, \bibinfo {author} {\bibnamefont {Jiménez-Escobar},
  \bibfnamefont {A.}}, \bibinfo {author} {\bibnamefont {Cecchi-Pestellini},
  \bibfnamefont {C.}}, \bibinfo {author} {\bibnamefont {Hsiao}, \bibfnamefont
  {L.-C.}}, \bibinfo {author} {\bibnamefont {Huang}, \bibfnamefont {C.-H.}}, \
  and\ \bibinfo {author} {\bibnamefont {Chen}, \bibfnamefont {Y.-J.}},\
  }\bibfield  {title} {\enquote {\bibinfo {title} {X-ray processing of a
  realistic ice mantle can explain the gas abundances in protoplanetary
  disks},}\ }\href {\doibase 10.1073/pnas.2005225117} {\bibfield  {journal}
  {\bibinfo  {journal} {Proceedings of the National Academy of Sciences}\ ,\
  \bibinfo {pages} {202005225}} (\bibinfo {year} {2020})}\BibitemShut {NoStop}%
\bibitem [{\citenamefont {Cruikshank}\ \emph {et~al.}(1993)\citenamefont
  {Cruikshank}, \citenamefont {Roush}, \citenamefont {Owen}, \citenamefont
  {Geballe}, \citenamefont {de~Bergh}, \citenamefont {Schmitt}, \citenamefont
  {Brown},\ and\ \citenamefont {Bartholomew}}]{Cruikshank_1993}%
  \BibitemOpen
  \bibfield  {author} {\bibinfo {author} {\bibnamefont {Cruikshank},
  \bibfnamefont {D.~P.}}, \bibinfo {author} {\bibnamefont {Roush},
  \bibfnamefont {T.~L.}}, \bibinfo {author} {\bibnamefont {Owen}, \bibfnamefont
  {T.~C.}}, \bibinfo {author} {\bibnamefont {Geballe}, \bibfnamefont {T.~R.}},
  \bibinfo {author} {\bibnamefont {de~Bergh}, \bibfnamefont {C.}}, \bibinfo
  {author} {\bibnamefont {Schmitt}, \bibfnamefont {B.}}, \bibinfo {author}
  {\bibnamefont {Brown}, \bibfnamefont {R.~H.}}, \ and\ \bibinfo {author}
  {\bibnamefont {Bartholomew}, \bibfnamefont {M.~J.}},\ }\bibfield  {title}
  {\enquote {\bibinfo {title} {Ices on the surface of triton},}\ }\href
  {\doibase 10.1126/science.261.5122.742} {\bibfield  {journal} {\bibinfo
  {journal} {Science}\ }\textbf {\bibinfo {volume} {261}},\ \bibinfo {pages}
  {742--745} (\bibinfo {year} {1993})}\BibitemShut {NoStop}%
\bibitem [{\citenamefont {Doronin}\ \emph {et~al.}(2015)\citenamefont
  {Doronin}, \citenamefont {Bertin}, \citenamefont {Michaut}, \citenamefont
  {Philippe},\ and\ \citenamefont {Fillion}}]{doronin_adsorption_2015}%
  \BibitemOpen
  \bibfield  {author} {\bibinfo {author} {\bibnamefont {Doronin}, \bibfnamefont
  {M.}}, \bibinfo {author} {\bibnamefont {Bertin}, \bibfnamefont {M.}},
  \bibinfo {author} {\bibnamefont {Michaut}, \bibfnamefont {X.}}, \bibinfo
  {author} {\bibnamefont {Philippe}, \bibfnamefont {L.}}, \ and\ \bibinfo
  {author} {\bibnamefont {Fillion}, \bibfnamefont {J.-H.}},\ }\bibfield
  {title} {\enquote {\bibinfo {title} {Adsorption energies and prefactor
  determination for {CH} $_{\textrm{3}}$ {OH} adsorption on graphite},}\ }\href
  {\doibase 10.1063/1.4929376} {\bibfield  {journal} {\bibinfo  {journal} {The
  Journal of Chemical Physics}\ }\textbf {\bibinfo {volume} {143}},\ \bibinfo
  {pages} {084703} (\bibinfo {year} {2015})}\BibitemShut {NoStop}%
\bibitem [{\citenamefont {Dupuy}\ \emph
  {et~al.}(2021{\natexlab{a}})\citenamefont {Dupuy}, \citenamefont {Bertin},
  \citenamefont {F\'eraud}, \citenamefont {Michaut}, \citenamefont
  {Marie-Jeanne}, \citenamefont {Jeseck}, \citenamefont {Philippe},
  \citenamefont {Baglin}, \citenamefont {Cimino}, \citenamefont {Romanzin},\
  and\ \citenamefont {Fillion}}]{Dupuy_water_ind_2021}%
  \BibitemOpen
  \bibfield  {author} {\bibinfo {author} {\bibnamefont {Dupuy}, \bibfnamefont
  {R.}}, \bibinfo {author} {\bibnamefont {Bertin}, \bibfnamefont {M.}},
  \bibinfo {author} {\bibnamefont {F\'eraud}, \bibfnamefont {G.}}, \bibinfo
  {author} {\bibnamefont {Michaut}, \bibfnamefont {X.}}, \bibinfo {author}
  {\bibnamefont {Marie-Jeanne}, \bibfnamefont {P.}}, \bibinfo {author}
  {\bibnamefont {Jeseck}, \bibfnamefont {P.}}, \bibinfo {author} {\bibnamefont
  {Philippe}, \bibfnamefont {L.}}, \bibinfo {author} {\bibnamefont {Baglin},
  \bibfnamefont {V.}}, \bibinfo {author} {\bibnamefont {Cimino}, \bibfnamefont
  {R.}}, \bibinfo {author} {\bibnamefont {Romanzin}, \bibfnamefont {C.}}, \
  and\ \bibinfo {author} {\bibnamefont {Fillion}, \bibfnamefont {J.-H.}},\
  }\bibfield  {title} {\enquote {\bibinfo {title} {Mechanism of indirect
  photon-induced desorption at the water ice surface},}\ }\href {\doibase
  10.1103/PhysRevLett.126.156001} {\bibfield  {journal} {\bibinfo  {journal}
  {Phys. Rev. Lett.}\ }\textbf {\bibinfo {volume} {126}},\ \bibinfo {pages}
  {156001} (\bibinfo {year} {2021}{\natexlab{a}})}\BibitemShut {NoStop}%
\bibitem [{\citenamefont {Dupuy}\ \emph {et~al.}(2018)\citenamefont {Dupuy},
  \citenamefont {Bertin}, \citenamefont {Féraud}, \citenamefont {Hassenfratz},
  \citenamefont {Michaut}, \citenamefont {Putaud}, \citenamefont {Philippe},
  \citenamefont {Jeseck}, \citenamefont {Angelucci}, \citenamefont {Cimino},
  \citenamefont {Baglin}, \citenamefont {Romanzin},\ and\ \citenamefont
  {Fillion}}]{dupuy_x-ray_2018}%
  \BibitemOpen
  \bibfield  {author} {\bibinfo {author} {\bibnamefont {Dupuy}, \bibfnamefont
  {R.}}, \bibinfo {author} {\bibnamefont {Bertin}, \bibfnamefont {M.}},
  \bibinfo {author} {\bibnamefont {Féraud}, \bibfnamefont {G.}}, \bibinfo
  {author} {\bibnamefont {Hassenfratz}, \bibfnamefont {M.}}, \bibinfo {author}
  {\bibnamefont {Michaut}, \bibfnamefont {X.}}, \bibinfo {author} {\bibnamefont
  {Putaud}, \bibfnamefont {T.}}, \bibinfo {author} {\bibnamefont {Philippe},
  \bibfnamefont {L.}}, \bibinfo {author} {\bibnamefont {Jeseck}, \bibfnamefont
  {P.}}, \bibinfo {author} {\bibnamefont {Angelucci}, \bibfnamefont {M.}},
  \bibinfo {author} {\bibnamefont {Cimino}, \bibfnamefont {R.}}, \bibinfo
  {author} {\bibnamefont {Baglin}, \bibfnamefont {V.}}, \bibinfo {author}
  {\bibnamefont {Romanzin}, \bibfnamefont {C.}}, \ and\ \bibinfo {author}
  {\bibnamefont {Fillion}, \bibfnamefont {J.-H.}},\ }\bibfield  {title}
  {\enquote {\bibinfo {title} {X-ray photodesorption from water ice in
  protoplanetary disks and {X}-ray-dominated regions},}\ }\href {\doibase
  10.1038/s41550-018-0532-y} {\bibfield  {journal} {\bibinfo  {journal} {Nature
  Astronomy}\ }\textbf {\bibinfo {volume} {2}},\ \bibinfo {pages} {796--801}
  (\bibinfo {year} {2018})}\BibitemShut {NoStop}%
\bibitem [{\citenamefont {Dupuy}\ \emph
  {et~al.}(2021{\natexlab{b}})\citenamefont {Dupuy}, \citenamefont {Bertin},
  \citenamefont {Féraud}, \citenamefont {Romanzin}, \citenamefont {Putaud},
  \citenamefont {Philippe}, \citenamefont {Michaut}, \citenamefont {Jeseck},
  \citenamefont {Cimino}, \citenamefont {Baglin},\ and\ \citenamefont
  {Fillion}}]{dupuy_co_2021}%
  \BibitemOpen
  \bibfield  {author} {\bibinfo {author} {\bibnamefont {Dupuy}, \bibfnamefont
  {R.}}, \bibinfo {author} {\bibnamefont {Bertin}, \bibfnamefont {M.}},
  \bibinfo {author} {\bibnamefont {Féraud}, \bibfnamefont {G.}}, \bibinfo
  {author} {\bibnamefont {Romanzin}, \bibfnamefont {C.}}, \bibinfo {author}
  {\bibnamefont {Putaud}, \bibfnamefont {T.}}, \bibinfo {author} {\bibnamefont
  {Philippe}, \bibfnamefont {L.}}, \bibinfo {author} {\bibnamefont {Michaut},
  \bibfnamefont {X.}}, \bibinfo {author} {\bibnamefont {Jeseck}, \bibfnamefont
  {P.}}, \bibinfo {author} {\bibnamefont {Cimino}, \bibfnamefont {R.}},
  \bibinfo {author} {\bibnamefont {Baglin}, \bibfnamefont {V.}}, \ and\
  \bibinfo {author} {\bibnamefont {Fillion}, \bibfnamefont {J.-H.}},\
  }\bibfield  {title} {\enquote {\bibinfo {title} {X-ray induced desorption and
  photochemistry in co ice},}\ }\href {\doibase 10.1039/D1CP02670D} {\bibfield
  {journal} {\bibinfo  {journal} {Phys. Chem. Chem. Phys.}\ }\textbf {\bibinfo
  {volume} {23}},\ \bibinfo {pages} {15965--15979} (\bibinfo {year}
  {2021}{\natexlab{b}})}\BibitemShut {NoStop}%
\bibitem [{\citenamefont {Dupuy}\ \emph {et~al.}(2020)\citenamefont {Dupuy},
  \citenamefont {Féraud}, \citenamefont {Bertin}, \citenamefont {Romanzin},
  \citenamefont {Philippe}, \citenamefont {Putaud}, \citenamefont {Michaut},
  \citenamefont {Cimino}, \citenamefont {Baglin},\ and\ \citenamefont
  {Fillion}}]{dupuy_desorption_2020}%
  \BibitemOpen
  \bibfield  {author} {\bibinfo {author} {\bibnamefont {Dupuy}, \bibfnamefont
  {R.}}, \bibinfo {author} {\bibnamefont {Féraud}, \bibfnamefont {G.}},
  \bibinfo {author} {\bibnamefont {Bertin}, \bibfnamefont {M.}}, \bibinfo
  {author} {\bibnamefont {Romanzin}, \bibfnamefont {C.}}, \bibinfo {author}
  {\bibnamefont {Philippe}, \bibfnamefont {L.}}, \bibinfo {author}
  {\bibnamefont {Putaud}, \bibfnamefont {T.}}, \bibinfo {author} {\bibnamefont
  {Michaut}, \bibfnamefont {X.}}, \bibinfo {author} {\bibnamefont {Cimino},
  \bibfnamefont {R.}}, \bibinfo {author} {\bibnamefont {Baglin}, \bibfnamefont
  {V.}}, \ and\ \bibinfo {author} {\bibnamefont {Fillion}, \bibfnamefont
  {J.-H.}},\ }\bibfield  {title} {\enquote {\bibinfo {title} {Desorption of
  neutrals, cations, and anions from core-excited amorphous solid water},}\
  }\href {\doibase 10.1063/1.5133156} {\bibfield  {journal} {\bibinfo
  {journal} {The Journal of Chemical Physics}\ }\textbf {\bibinfo {volume}
  {152}},\ \bibinfo {pages} {054711} (\bibinfo {year} {2020})}\BibitemShut
  {NoStop}%
\bibitem [{\citenamefont {Fayolle}\ \emph {et~al.}(2013)\citenamefont
  {Fayolle}, \citenamefont {Bertin}, \citenamefont {Romanzin}, \citenamefont
  {M~Poderoso}, \citenamefont {Philippe}, \citenamefont {Michaut},
  \citenamefont {Jeseck}, \citenamefont {Linnartz}, \citenamefont {Öberg},\
  and\ \citenamefont {Fillion}}]{fayolle_wavelength-dependent_2013}%
  \BibitemOpen
  \bibfield  {author} {\bibinfo {author} {\bibnamefont {Fayolle}, \bibfnamefont
  {E.~C.}}, \bibinfo {author} {\bibnamefont {Bertin}, \bibfnamefont {M.}},
  \bibinfo {author} {\bibnamefont {Romanzin}, \bibfnamefont {C.}}, \bibinfo
  {author} {\bibnamefont {M~Poderoso}, \bibfnamefont {H.~A.}}, \bibinfo
  {author} {\bibnamefont {Philippe}, \bibfnamefont {L.}}, \bibinfo {author}
  {\bibnamefont {Michaut}, \bibfnamefont {X.}}, \bibinfo {author} {\bibnamefont
  {Jeseck}, \bibfnamefont {P.}}, \bibinfo {author} {\bibnamefont {Linnartz},
  \bibfnamefont {H.}}, \bibinfo {author} {\bibnamefont {Öberg}, \bibfnamefont
  {K.~I.}}, \ and\ \bibinfo {author} {\bibnamefont {Fillion}, \bibfnamefont
  {J.-H.}},\ }\bibfield  {title} {\enquote {\bibinfo {title}
  {Wavelength-dependent {UV} photodesorption of pure {N} $_{\textrm{2}}$ and
  {O} $_{\textrm{2}}$ ices},}\ }\href {\doibase 10.1051/0004-6361/201321533}
  {\bibfield  {journal} {\bibinfo  {journal} {Astronomy \& Astrophysics}\
  }\textbf {\bibinfo {volume} {556}},\ \bibinfo {pages} {A122} (\bibinfo {year}
  {2013})}\BibitemShut {NoStop}%
\bibitem [{\citenamefont {Fayolle}\ \emph {et~al.}(2011)\citenamefont
  {Fayolle}, \citenamefont {Bertin}, \citenamefont {Romanzin}, \citenamefont
  {Michaut}, \citenamefont {Öberg}, \citenamefont {Linnartz},\ and\
  \citenamefont {Fillion}}]{fayolle_co_2011}%
  \BibitemOpen
  \bibfield  {author} {\bibinfo {author} {\bibnamefont {Fayolle}, \bibfnamefont
  {E.~C.}}, \bibinfo {author} {\bibnamefont {Bertin}, \bibfnamefont {M.}},
  \bibinfo {author} {\bibnamefont {Romanzin}, \bibfnamefont {C.}}, \bibinfo
  {author} {\bibnamefont {Michaut}, \bibfnamefont {X.}}, \bibinfo {author}
  {\bibnamefont {Öberg}, \bibfnamefont {K.~I.}}, \bibinfo {author}
  {\bibnamefont {Linnartz}, \bibfnamefont {H.}}, \ and\ \bibinfo {author}
  {\bibnamefont {Fillion}, \bibfnamefont {J.-H.}},\ }\bibfield  {title}
  {\enquote {\bibinfo {title} {{CO} {ICE} {PHOTODESORPTION}: {A}
  {WAVELENGTH}-{DEPENDENT} {STUDY}},}\ }\href {\doibase
  10.1088/2041-8205/739/2/L36} {\bibfield  {journal} {\bibinfo  {journal} {The
  Astrophysical Journal}\ }\textbf {\bibinfo {volume} {739}},\ \bibinfo {pages}
  {L36} (\bibinfo {year} {2011})}\BibitemShut {NoStop}%
\bibitem [{\citenamefont {Feifel}\ \emph {et~al.}(2004)\citenamefont {Feifel},
  \citenamefont {Andersson}, \citenamefont {Öhrwall}, \citenamefont
  {Sorensen}, \citenamefont {Piancastelli}, \citenamefont {Tchaplyguine},
  \citenamefont {Björneholm}, \citenamefont {Karlsson},\ and\ \citenamefont
  {Svensson}}]{feifel_quantitative_2004}%
  \BibitemOpen
  \bibfield  {author} {\bibinfo {author} {\bibnamefont {Feifel}, \bibfnamefont
  {R.}}, \bibinfo {author} {\bibnamefont {Andersson}, \bibfnamefont {M.}},
  \bibinfo {author} {\bibnamefont {Öhrwall}, \bibfnamefont {G.}}, \bibinfo
  {author} {\bibnamefont {Sorensen}, \bibfnamefont {S.}}, \bibinfo {author}
  {\bibnamefont {Piancastelli}, \bibfnamefont {M.}}, \bibinfo {author}
  {\bibnamefont {Tchaplyguine}, \bibfnamefont {M.}}, \bibinfo {author}
  {\bibnamefont {Björneholm}, \bibfnamefont {O.}}, \bibinfo {author}
  {\bibnamefont {Karlsson}, \bibfnamefont {L.}}, \ and\ \bibinfo {author}
  {\bibnamefont {Svensson}, \bibfnamefont {S.}},\ }\bibfield  {title} {\enquote
  {\bibinfo {title} {A quantitative analysis of the {N} 1s→pi photoabsorption
  profile in {N2}: new spectroscopical constants for the core-excited state},}\
  }\href {\doibase 10.1016/j.cplett.2003.11.026} {\bibfield  {journal}
  {\bibinfo  {journal} {Chemical Physics Letters}\ }\textbf {\bibinfo {volume}
  {383}},\ \bibinfo {pages} {222--229} (\bibinfo {year} {2004})}\BibitemShut
  {NoStop}%
\bibitem [{\citenamefont {Feulner}\ \emph {et~al.}(2002)\citenamefont
  {Feulner}, \citenamefont {Ecker}, \citenamefont {Romberg}, \citenamefont
  {Weimar},\ and\ \citenamefont
  {Föhlisch}}]{feulner_core-excitation-induced_2002}%
  \BibitemOpen
  \bibfield  {author} {\bibinfo {author} {\bibnamefont {Feulner}, \bibfnamefont
  {P.}}, \bibinfo {author} {\bibnamefont {Ecker}, \bibfnamefont {M.}}, \bibinfo
  {author} {\bibnamefont {Romberg}, \bibfnamefont {R.}}, \bibinfo {author}
  {\bibnamefont {Weimar}, \bibfnamefont {R.}}, \ and\ \bibinfo {author}
  {\bibnamefont {Föhlisch}, \bibfnamefont {A.}},\ }\bibfield  {title}
  {\enquote {\bibinfo {title} {{CORE}-{EXCITATION}-{INDUCED} {BOND} {BREAKING}
  {OF} {CHEMISORBED} {MOLECULES} {PROBED} {BY} {EMISSION} {OF} {IONS},
  {NEUTRALS}, {AND} {ELECTRONS}},}\ }\href {\doibase 10.1142/S0218625X02001859}
  {\bibfield  {journal} {\bibinfo  {journal} {Surface Review and Letters}\
  }\textbf {\bibinfo {volume} {09}},\ \bibinfo {pages} {759--768} (\bibinfo
  {year} {2002})}\BibitemShut {NoStop}%
\bibitem [{\citenamefont {Feulner}\ \emph {et~al.}(2000)\citenamefont
  {Feulner}, \citenamefont {Romberg}, \citenamefont {Frigo}, \citenamefont
  {Weimar}, \citenamefont {Gsell}, \citenamefont {Ogurtsov},\ and\
  \citenamefont {Menzel}}]{feulner_recent_2000}%
  \BibitemOpen
  \bibfield  {author} {\bibinfo {author} {\bibnamefont {Feulner}, \bibfnamefont
  {P.}}, \bibinfo {author} {\bibnamefont {Romberg}, \bibfnamefont {R.}},
  \bibinfo {author} {\bibnamefont {Frigo}, \bibfnamefont {S.}}, \bibinfo
  {author} {\bibnamefont {Weimar}, \bibfnamefont {R.}}, \bibinfo {author}
  {\bibnamefont {Gsell}, \bibfnamefont {M.}}, \bibinfo {author} {\bibnamefont
  {Ogurtsov}, \bibfnamefont {A.}}, \ and\ \bibinfo {author} {\bibnamefont
  {Menzel}, \bibfnamefont {D.}},\ }\bibfield  {title} {\enquote {\bibinfo
  {title} {Recent progress in the investigation of core hole-induced photon
  stimulated desorption from adsorbates: excitation site-dependent bond
  breaking, and charge rearrangement},}\ }\href {\doibase
  10.1016/S0039-6028(00)00006-6} {\bibfield  {journal} {\bibinfo  {journal}
  {Surface Science}\ }\textbf {\bibinfo {volume} {451}},\ \bibinfo {pages}
  {41--52} (\bibinfo {year} {2000})}\BibitemShut {NoStop}%
\bibitem [{\citenamefont {Feulner}\ \emph {et~al.}(1992)\citenamefont
  {Feulner}, \citenamefont {Scheuerer}, \citenamefont {Scheuer}, \citenamefont
  {Remmers}, \citenamefont {Wurth},\ and\ \citenamefont
  {Menzel}}]{feulner_high_1992}%
  \BibitemOpen
  \bibfield  {author} {\bibinfo {author} {\bibnamefont {Feulner}, \bibfnamefont
  {P.}}, \bibinfo {author} {\bibnamefont {Scheuerer}, \bibfnamefont {R.}},
  \bibinfo {author} {\bibnamefont {Scheuer}, \bibfnamefont {M.}}, \bibinfo
  {author} {\bibnamefont {Remmers}, \bibfnamefont {G.}}, \bibinfo {author}
  {\bibnamefont {Wurth}, \bibfnamefont {W.}}, \ and\ \bibinfo {author}
  {\bibnamefont {Menzel}, \bibfnamefont {D.}},\ }\bibfield  {title} {\enquote
  {\bibinfo {title} {High resolution photon stimulated desorption spectroscopy
  of solid nitrogen by resonant {N} 1s core level excitation},}\ }\href
  {\doibase 10.1007/BF00348336} {\bibfield  {journal} {\bibinfo  {journal}
  {Applied Physics A}\ }\textbf {\bibinfo {volume} {55}},\ \bibinfo {pages}
  {478--481} (\bibinfo {year} {1992})}\BibitemShut {NoStop}%
\bibitem [{\citenamefont {Frigo}\ \emph {et~al.}(1998)\citenamefont {Frigo},
  \citenamefont {Feulner}, \citenamefont {Kassühlke}, \citenamefont {Keller},\
  and\ \citenamefont {Menzel}}]{frigo_observation_1998}%
  \BibitemOpen
  \bibfield  {author} {\bibinfo {author} {\bibnamefont {Frigo}, \bibfnamefont
  {S.~P.}}, \bibinfo {author} {\bibnamefont {Feulner}, \bibfnamefont {P.}},
  \bibinfo {author} {\bibnamefont {Kassühlke}, \bibfnamefont {B.}}, \bibinfo
  {author} {\bibnamefont {Keller}, \bibfnamefont {C.}}, \ and\ \bibinfo
  {author} {\bibnamefont {Menzel}, \bibfnamefont {D.}},\ }\bibfield  {title}
  {\enquote {\bibinfo {title} {Observation of {Neutral} {Atomic} {Fragments}
  for {Specific} 1 s {Core} {Excitations} of an {Adsorbed} {Molecule}},}\
  }\href {\doibase 10.1103/PhysRevLett.80.2813} {\bibfield  {journal} {\bibinfo
   {journal} {Physical Review Letters}\ }\textbf {\bibinfo {volume} {80}},\
  \bibinfo {pages} {2813--2816} (\bibinfo {year} {1998})}\BibitemShut {NoStop}%
\bibitem [{\citenamefont {Hitchcock}\ and\ \citenamefont
  {Brion}(1980)}]{HITCHCOCK_1980}%
  \BibitemOpen
  \bibfield  {author} {\bibinfo {author} {\bibnamefont {Hitchcock},
  \bibfnamefont {A.}}\ and\ \bibinfo {author} {\bibnamefont {Brion},
  \bibfnamefont {C.}},\ }\bibfield  {title} {\enquote {\bibinfo {title}
  {K-shell excitation spectra of co, n2 and o2},}\ }\href {\doibase
  https://doi.org/10.1016/0368-2048(80)80001-6} {\bibfield  {journal} {\bibinfo
   {journal} {Journal of Electron Spectroscopy and Related Phenomena}\ }\textbf
  {\bibinfo {volume} {18}},\ \bibinfo {pages} {1--21} (\bibinfo {year}
  {1980})}\BibitemShut {NoStop}%
\bibitem [{\citenamefont {Jiménez-Escobar}\ \emph {et~al.}(2018)\citenamefont
  {Jiménez-Escobar}, \citenamefont {Ciaravella}, \citenamefont
  {Cecchi-Pestellini}, \citenamefont {Huang}, \citenamefont {Sie},
  \citenamefont {Chen},\ and\ \citenamefont
  {Muñoz~Caro}}]{jimenez-escobar_x-ray_2018}%
  \BibitemOpen
  \bibfield  {author} {\bibinfo {author} {\bibnamefont {Jiménez-Escobar},
  \bibfnamefont {A.}}, \bibinfo {author} {\bibnamefont {Ciaravella},
  \bibfnamefont {A.}}, \bibinfo {author} {\bibnamefont {Cecchi-Pestellini},
  \bibfnamefont {C.}}, \bibinfo {author} {\bibnamefont {Huang}, \bibfnamefont
  {C.-H.}}, \bibinfo {author} {\bibnamefont {Sie}, \bibfnamefont {N.-E.}},
  \bibinfo {author} {\bibnamefont {Chen}, \bibfnamefont {Y.-J.}}, \ and\
  \bibinfo {author} {\bibnamefont {Muñoz~Caro}, \bibfnamefont {G.~M.}},\
  }\bibfield  {title} {\enquote {\bibinfo {title} {X-{Ray} {Photo}-desorption
  of {H} $_{\textrm{2}}$ {O}:{CO}:{NH} $_{\textrm{3}}$ {Circumstellar} {Ice}
  {Analogs}: {Gas}-phase {Enrichment}},}\ }\href {\doibase
  10.3847/1538-4357/aae711} {\bibfield  {journal} {\bibinfo  {journal} {The
  Astrophysical Journal}\ }\textbf {\bibinfo {volume} {868}},\ \bibinfo {pages}
  {73} (\bibinfo {year} {2018})}\BibitemShut {NoStop}%
\bibitem [{\citenamefont {Jugnet}\ \emph {et~al.}(1984)\citenamefont {Jugnet},
  \citenamefont {Himpsel}, \citenamefont {Avouris},\ and\ \citenamefont
  {Koch}}]{jugnet_high-resolution_1984}%
  \BibitemOpen
  \bibfield  {author} {\bibinfo {author} {\bibnamefont {Jugnet}, \bibfnamefont
  {Y.}}, \bibinfo {author} {\bibnamefont {Himpsel}, \bibfnamefont {F.~J.}},
  \bibinfo {author} {\bibnamefont {Avouris}, \bibfnamefont {P.}}, \ and\
  \bibinfo {author} {\bibnamefont {Koch}, \bibfnamefont {E.~E.}},\ }\bibfield
  {title} {\enquote {\bibinfo {title} {High-{Resolution} {C} 1 s and {O} 1 s
  {Core}-{Excitation} {Spectroscopy} of {Chemisorbed}, {Physisorbed}, and
  {Free} {CO}},}\ }\href {\doibase 10.1103/PhysRevLett.53.198} {\bibfield
  {journal} {\bibinfo  {journal} {Physical Review Letters}\ }\textbf {\bibinfo
  {volume} {53}},\ \bibinfo {pages} {198--201} (\bibinfo {year}
  {1984})}\BibitemShut {NoStop}%
\bibitem [{\citenamefont {Kato}\ \emph {et~al.}(2007)\citenamefont {Kato},
  \citenamefont {Morishita}, \citenamefont {Oura}, \citenamefont {Yamaoka},
  \citenamefont {Tamenori}, \citenamefont {Okada}, \citenamefont {Matsudo},
  \citenamefont {Gejo}, \citenamefont {Suzuki},\ and\ \citenamefont
  {Saito}}]{kato_absolute_2007}%
  \BibitemOpen
  \bibfield  {author} {\bibinfo {author} {\bibnamefont {Kato}, \bibfnamefont
  {M.}}, \bibinfo {author} {\bibnamefont {Morishita}, \bibfnamefont {Y.}},
  \bibinfo {author} {\bibnamefont {Oura}, \bibfnamefont {M.}}, \bibinfo
  {author} {\bibnamefont {Yamaoka}, \bibfnamefont {H.}}, \bibinfo {author}
  {\bibnamefont {Tamenori}, \bibfnamefont {Y.}}, \bibinfo {author}
  {\bibnamefont {Okada}, \bibfnamefont {K.}}, \bibinfo {author} {\bibnamefont
  {Matsudo}, \bibfnamefont {T.}}, \bibinfo {author} {\bibnamefont {Gejo},
  \bibfnamefont {T.}}, \bibinfo {author} {\bibnamefont {Suzuki}, \bibfnamefont
  {I.}}, \ and\ \bibinfo {author} {\bibnamefont {Saito}, \bibfnamefont {N.}},\
  }\bibfield  {title} {\enquote {\bibinfo {title} {Absolute photoionization
  cross sections with ultra-high energy resolution for {Ar}, {Kr}, {Xe} and
  {N2} in inner-shell ionization regions},}\ }\href {\doibase
  10.1016/j.elspec.2007.06.003} {\bibfield  {journal} {\bibinfo  {journal}
  {Journal of Electron Spectroscopy and Related Phenomena}\ }\textbf {\bibinfo
  {volume} {160}},\ \bibinfo {pages} {39--48} (\bibinfo {year}
  {2007})}\BibitemShut {NoStop}%
\bibitem [{\citenamefont {Kempgens}\ \emph {et~al.}(1996)\citenamefont
  {Kempgens}, \citenamefont {Kivimäki}, \citenamefont {Neeb}, \citenamefont
  {Köppe}, \citenamefont {Bradshaw},\ and\ \citenamefont
  {Feldhaus}}]{Kempgens_1996}%
  \BibitemOpen
  \bibfield  {author} {\bibinfo {author} {\bibnamefont {Kempgens},
  \bibfnamefont {B.}}, \bibinfo {author} {\bibnamefont {Kivimäki},
  \bibfnamefont {A.}}, \bibinfo {author} {\bibnamefont {Neeb}, \bibfnamefont
  {M.}}, \bibinfo {author} {\bibnamefont {Köppe}, \bibfnamefont {H.~M.}},
  \bibinfo {author} {\bibnamefont {Bradshaw}, \bibfnamefont {A.~M.}}, \ and\
  \bibinfo {author} {\bibnamefont {Feldhaus}, \bibfnamefont {J.}},\ }\bibfield
  {title} {\enquote {\bibinfo {title} {A high-resolution n 1s photoionization
  study of the molecule in the near-threshold region},}\ }\href {\doibase
  10.1088/0953-4075/29/22/016} {\bibfield  {journal} {\bibinfo  {journal}
  {Journal of Physics B: Atomic, Molecular and Optical Physics}\ }\textbf
  {\bibinfo {volume} {29}},\ \bibinfo {pages} {5389--5402} (\bibinfo {year}
  {1996})}\BibitemShut {NoStop}%
\bibitem [{\citenamefont {King}, \citenamefont {Read},\ and\ \citenamefont
  {Tronc}(1977)}]{KING_1977}%
  \BibitemOpen
  \bibfield  {author} {\bibinfo {author} {\bibnamefont {King}, \bibfnamefont
  {G.~C.}}, \bibinfo {author} {\bibnamefont {Read}, \bibfnamefont {F.~H.}}, \
  and\ \bibinfo {author} {\bibnamefont {Tronc}, \bibfnamefont {M.}},\
  }\bibfield  {title} {\enquote {\bibinfo {title} {Investigation of the energy
  and vibrational structure of the inner shell (1s)->1(pi2p)1pi state of the
  nitrogen molecule by electron impact with high resolution},}\ }\href
  {\doibase https://doi.org/10.1016/0009-2614(77)85118-X} {\bibfield  {journal}
  {\bibinfo  {journal} {Chemical Physics Letters}\ }\textbf {\bibinfo {volume}
  {52}},\ \bibinfo {pages} {50--54} (\bibinfo {year} {1977})}\BibitemShut
  {NoStop}%
\bibitem [{\citenamefont {Krause}(1979)}]{Krause_1979}%
  \BibitemOpen
  \bibfield  {author} {\bibinfo {author} {\bibnamefont {Krause}, \bibfnamefont
  {M.~O.}},\ }\bibfield  {title} {\enquote {\bibinfo {title} {Atomic radiative
  and radiationless yields for k and l shells},}\ }\href {\doibase
  10.1063/1.555594} {\bibfield  {journal} {\bibinfo  {journal} {Journal of
  Physical and Chemical Reference Data}\ }\textbf {\bibinfo {volume} {8}},\
  \bibinfo {pages} {307--327} (\bibinfo {year} {1979})}\BibitemShut {NoStop}%
\bibitem [{\citenamefont {Marchione}\ and\ \citenamefont
  {McCoustra}(2017)}]{marchione_electron-induced_2017}%
  \BibitemOpen
  \bibfield  {author} {\bibinfo {author} {\bibnamefont {Marchione},
  \bibfnamefont {D.}}\ and\ \bibinfo {author} {\bibnamefont {McCoustra},
  \bibfnamefont {M.~R.~S.}},\ }\bibfield  {title} {\enquote {\bibinfo {title}
  {Electron-{Induced} {Chemistry}: {Preliminary} {Comparative} {Studies} of
  {Hydrogen} {Production} from {Water}, {Methanol}, and {Diethyl} {Ether}},}\
  }\href {\doibase 10.1021/acsearthspacechem.7b00032} {\bibfield  {journal}
  {\bibinfo  {journal} {ACS Earth and Space Chemistry}\ }\textbf {\bibinfo
  {volume} {1}},\ \bibinfo {pages} {310--315} (\bibinfo {year}
  {2017})}\BibitemShut {NoStop}%
\bibitem [{\citenamefont {Marchione}, \citenamefont {Thrower},\ and\
  \citenamefont {McCoustra}(2016)}]{marchione_efficient_2016}%
  \BibitemOpen
  \bibfield  {author} {\bibinfo {author} {\bibnamefont {Marchione},
  \bibfnamefont {D.}}, \bibinfo {author} {\bibnamefont {Thrower}, \bibfnamefont
  {J.~D.}}, \ and\ \bibinfo {author} {\bibnamefont {McCoustra}, \bibfnamefont
  {M.~R.~S.}},\ }\bibfield  {title} {\enquote {\bibinfo {title} {Efficient
  electron-promoted desorption of benzene from water ice surfaces},}\ }\href
  {\doibase 10.1039/C5CP06537B} {\bibfield  {journal} {\bibinfo  {journal}
  {Physical Chemistry Chemical Physics}\ }\textbf {\bibinfo {volume} {18}},\
  \bibinfo {pages} {4026--4034} (\bibinfo {year} {2016})}\BibitemShut {NoStop}%
\bibitem [{\citenamefont {Moddeman}\ \emph {et~al.}(1971)\citenamefont
  {Moddeman}, \citenamefont {Carlson}, \citenamefont {Krause}, \citenamefont
  {Pullen}, \citenamefont {Bull},\ and\ \citenamefont
  {Schweitzer}}]{moddeman_determination_1971}%
  \BibitemOpen
  \bibfield  {author} {\bibinfo {author} {\bibnamefont {Moddeman},
  \bibfnamefont {W.~E.}}, \bibinfo {author} {\bibnamefont {Carlson},
  \bibfnamefont {T.~A.}}, \bibinfo {author} {\bibnamefont {Krause},
  \bibfnamefont {M.~O.}}, \bibinfo {author} {\bibnamefont {Pullen},
  \bibfnamefont {B.~P.}}, \bibinfo {author} {\bibnamefont {Bull}, \bibfnamefont
  {W.~E.}}, \ and\ \bibinfo {author} {\bibnamefont {Schweitzer}, \bibfnamefont
  {G.~K.}},\ }\bibfield  {title} {\enquote {\bibinfo {title} {Determination of
  the \textit{{K}—{LL}} {Auger} {Spectra} of {N} $_{\textrm{2}}$ , {O}
  $_{\textrm{2}}$ , {CO}, {NO}, {H} $_{\textrm{2}}$ {O}, and {CO}
  $_{\textrm{2}}$},}\ }\href {\doibase 10.1063/1.1676411} {\bibfield  {journal}
  {\bibinfo  {journal} {The Journal of Chemical Physics}\ }\textbf {\bibinfo
  {volume} {55}},\ \bibinfo {pages} {2317--2336} (\bibinfo {year}
  {1971})}\BibitemShut {NoStop}%
\bibitem [{\citenamefont {{Mu\~noz Caro, G. M.}}\ \emph
  {et~al.}(2010)\citenamefont {{Mu\~noz Caro, G. M.}}, \citenamefont
  {{Jim\'enez-Escobar, A.}}, \citenamefont {{Mart\'{\i}n-Gago, J. \'A.}},
  \citenamefont {{Rogero, C.}}, \citenamefont {{Atienza, C.}}, \citenamefont
  {{Puertas, S.}}, \citenamefont {{Sobrado, J. M.}},\ and\ \citenamefont
  {{Torres-Redondo, J.}}}]{munoz_caro_new_2010}%
  \BibitemOpen
  \bibfield  {author} {\bibinfo {author} {\bibnamefont {{Mu\~noz Caro, G.
  M.}},}, \bibinfo {author} {\bibnamefont {{Jim\'enez-Escobar, A.}},}, \bibinfo
  {author} {\bibnamefont {{Mart\'{\i}n-Gago, J. \'A.}},}, \bibinfo {author}
  {\bibnamefont {{Rogero, C.}},}, \bibinfo {author} {\bibnamefont {{Atienza,
  C.}},}, \bibinfo {author} {\bibnamefont {{Puertas, S.}},}, \bibinfo {author}
  {\bibnamefont {{Sobrado, J. M.}},}, \ and\ \bibinfo {author} {\bibnamefont
  {{Torres-Redondo, J.}},},\ }\bibfield  {title} {\enquote {\bibinfo {title}
  {New results on thermal and photodesorption of co ice using the novel
  interstellar astrochemistry chamber (isac)},}\ }\href {\doibase
  10.1051/0004-6361/200912462} {\bibfield  {journal} {\bibinfo  {journal}
  {A\&A}\ }\textbf {\bibinfo {volume} {522}},\ \bibinfo {pages} {A108}
  (\bibinfo {year} {2010})}\BibitemShut {NoStop}%
\bibitem [{\citenamefont {Mumma}\ and\ \citenamefont
  {Charnley}(2011)}]{Mumma_2011}%
  \BibitemOpen
  \bibfield  {author} {\bibinfo {author} {\bibnamefont {Mumma}, \bibfnamefont
  {M.~J.}}\ and\ \bibinfo {author} {\bibnamefont {Charnley}, \bibfnamefont
  {S.~B.}},\ }\bibfield  {title} {\enquote {\bibinfo {title} {The chemical
  composition of comets—emerging taxonomies and natal heritage},}\ }\href
  {\doibase 10.1146/annurev-astro-081309-130811} {\bibfield  {journal}
  {\bibinfo  {journal} {Annual Review of Astronomy and Astrophysics}\ }\textbf
  {\bibinfo {volume} {49}},\ \bibinfo {pages} {471--524} (\bibinfo {year}
  {2011})}\BibitemShut {NoStop}%
\bibitem [{\citenamefont {Orlando}\ and\ \citenamefont
  {Kimmel}(1997)}]{Orlando_1997}%
  \BibitemOpen
  \bibfield  {author} {\bibinfo {author} {\bibnamefont {Orlando}, \bibfnamefont
  {T.}}\ and\ \bibinfo {author} {\bibnamefont {Kimmel}, \bibfnamefont {G.}},\
  }\bibfield  {title} {\enquote {\bibinfo {title} {The role of excitons and
  substrate temperature in low-energy (5–50 ev) electron-stimulated
  dissociation of amorphous d2o ice},}\ }\href {\doibase
  https://doi.org/10.1016/S0039-6028(97)00511-6} {\bibfield  {journal}
  {\bibinfo  {journal} {Surface Science}\ }\textbf {\bibinfo {volume} {390}},\
  \bibinfo {pages} {79--85} (\bibinfo {year} {1997})}\BibitemShut {NoStop}%
\bibitem [{\citenamefont {Owen}\ \emph {et~al.}(1993)\citenamefont {Owen},
  \citenamefont {Roush}, \citenamefont {Cruikshank}, \citenamefont {Elliot},
  \citenamefont {Young}, \citenamefont {de~Bergh}, \citenamefont {Schmitt},
  \citenamefont {Geballe}, \citenamefont {Brown},\ and\ \citenamefont
  {Bartholomew}}]{Owen_1993}%
  \BibitemOpen
  \bibfield  {author} {\bibinfo {author} {\bibnamefont {Owen}, \bibfnamefont
  {T.~C.}}, \bibinfo {author} {\bibnamefont {Roush}, \bibfnamefont {T.~L.}},
  \bibinfo {author} {\bibnamefont {Cruikshank}, \bibfnamefont {D.~P.}},
  \bibinfo {author} {\bibnamefont {Elliot}, \bibfnamefont {J.~L.}}, \bibinfo
  {author} {\bibnamefont {Young}, \bibfnamefont {L.~A.}}, \bibinfo {author}
  {\bibnamefont {de~Bergh}, \bibfnamefont {C.}}, \bibinfo {author}
  {\bibnamefont {Schmitt}, \bibfnamefont {B.}}, \bibinfo {author} {\bibnamefont
  {Geballe}, \bibfnamefont {T.~R.}}, \bibinfo {author} {\bibnamefont {Brown},
  \bibfnamefont {R.~H.}}, \ and\ \bibinfo {author} {\bibnamefont {Bartholomew},
  \bibfnamefont {M.~J.}},\ }\bibfield  {title} {\enquote {\bibinfo {title}
  {Surface ices and the atmospheric composition of pluto},}\ }\href {\doibase
  10.1126/science.261.5122.745} {\bibfield  {journal} {\bibinfo  {journal}
  {Science}\ }\textbf {\bibinfo {volume} {261}},\ \bibinfo {pages} {745--748}
  (\bibinfo {year} {1993})}\BibitemShut {NoStop}%
\bibitem [{\citenamefont {Petrik}\ and\ \citenamefont
  {Kimmel}(2003)}]{Petrik_2003}%
  \BibitemOpen
  \bibfield  {author} {\bibinfo {author} {\bibnamefont {Petrik}, \bibfnamefont
  {N.~G.}}\ and\ \bibinfo {author} {\bibnamefont {Kimmel}, \bibfnamefont
  {G.~A.}},\ }\bibfield  {title} {\enquote {\bibinfo {title}
  {Electron-stimulated reactions at the interfaces of amorphous solid water
  films driven by long-range energy transfer from the bulk},}\ }\href {\doibase
  10.1103/PhysRevLett.90.166102} {\bibfield  {journal} {\bibinfo  {journal}
  {Phys. Rev. Lett.}\ }\textbf {\bibinfo {volume} {90}},\ \bibinfo {pages}
  {166102} (\bibinfo {year} {2003})}\BibitemShut {NoStop}%
\bibitem [{\citenamefont {Rakhovskaia}, \citenamefont {Wiethoff},\ and\
  \citenamefont {Feulner}(1995)}]{RAKHOVSKAIA_1995}%
  \BibitemOpen
  \bibfield  {author} {\bibinfo {author} {\bibnamefont {Rakhovskaia},
  \bibfnamefont {O.}}, \bibinfo {author} {\bibnamefont {Wiethoff},
  \bibfnamefont {P.}}, \ and\ \bibinfo {author} {\bibnamefont {Feulner},
  \bibfnamefont {P.}},\ }\bibfield  {title} {\enquote {\bibinfo {title}
  {Thresholds for electron stimulated desorption of neutral molecules from
  solid n2, co, o2 and no},}\ }\href {\doibase
  https://doi.org/10.1016/0168-583X(95)00296-0} {\bibfield  {journal} {\bibinfo
   {journal} {Nuclear Instruments and Methods in Physics Research Section B:
  Beam Interactions with Materials and Atoms}\ }\textbf {\bibinfo {volume}
  {101}},\ \bibinfo {pages} {169--173} (\bibinfo {year} {1995})}\BibitemShut
  {NoStop}%
\bibitem [{\citenamefont {Romberg}\ \emph {et~al.}(2000)\citenamefont
  {Romberg}, \citenamefont {Heckmair}, \citenamefont {Frigo}, \citenamefont
  {Ogurtsov}, \citenamefont {Menzel},\ and\ \citenamefont
  {Feulner}}]{romberg_atom-selective_2000}%
  \BibitemOpen
  \bibfield  {author} {\bibinfo {author} {\bibnamefont {Romberg}, \bibfnamefont
  {R.}}, \bibinfo {author} {\bibnamefont {Heckmair}, \bibfnamefont {N.}},
  \bibinfo {author} {\bibnamefont {Frigo}, \bibfnamefont {S.~P.}}, \bibinfo
  {author} {\bibnamefont {Ogurtsov}, \bibfnamefont {A.}}, \bibinfo {author}
  {\bibnamefont {Menzel}, \bibfnamefont {D.}}, \ and\ \bibinfo {author}
  {\bibnamefont {Feulner}, \bibfnamefont {P.}},\ }\bibfield  {title} {\enquote
  {\bibinfo {title} {Atom-{Selective} {Bond} {Breaking} in a {Chemisorbed}
  {Homonuclear} {Molecule} {Induced} by {Core} {Excitation}: {N} 2 / {Ru} ( 001
  )},}\ }\href {\doibase 10.1103/PhysRevLett.84.374} {\bibfield  {journal}
  {\bibinfo  {journal} {Physical Review Letters}\ }\textbf {\bibinfo {volume}
  {84}},\ \bibinfo {pages} {374--377} (\bibinfo {year} {2000})}\BibitemShut
  {NoStop}%
\bibitem [{\citenamefont {Rubin}\ \emph {et~al.}(2015)\citenamefont {Rubin},
  \citenamefont {Altwegg}, \citenamefont {Balsiger}, \citenamefont {Bar-Nun},
  \citenamefont {Berthelier}, \citenamefont {Bieler}, \citenamefont {Bochsler},
  \citenamefont {Briois}, \citenamefont {Calmonte}, \citenamefont {Combi},
  \citenamefont {Keyser}, \citenamefont {Dhooghe}, \citenamefont {Eberhardt},
  \citenamefont {Fiethe}, \citenamefont {Fuselier}, \citenamefont {Gasc},
  \citenamefont {Gombosi}, \citenamefont {Hansen}, \citenamefont {Hässig},
  \citenamefont {Jäckel}, \citenamefont {Kopp}, \citenamefont {Korth},
  \citenamefont {Roy}, \citenamefont {Mall}, \citenamefont {Marty},
  \citenamefont {Mousis}, \citenamefont {Owen}, \citenamefont {Rème},
  \citenamefont {Sémon}, \citenamefont {Tzou}, \citenamefont {Waite},\ and\
  \citenamefont {Wurz}}]{Rubin_2015}%
  \BibitemOpen
  \bibfield  {author} {\bibinfo {author} {\bibnamefont {Rubin}, \bibfnamefont
  {M.}}, \bibinfo {author} {\bibnamefont {Altwegg}, \bibfnamefont {K.}},
  \bibinfo {author} {\bibnamefont {Balsiger}, \bibfnamefont {H.}}, \bibinfo
  {author} {\bibnamefont {Bar-Nun}, \bibfnamefont {A.}}, \bibinfo {author}
  {\bibnamefont {Berthelier}, \bibfnamefont {J.-J.}}, \bibinfo {author}
  {\bibnamefont {Bieler}, \bibfnamefont {A.}}, \bibinfo {author} {\bibnamefont
  {Bochsler}, \bibfnamefont {P.}}, \bibinfo {author} {\bibnamefont {Briois},
  \bibfnamefont {C.}}, \bibinfo {author} {\bibnamefont {Calmonte},
  \bibfnamefont {U.}}, \bibinfo {author} {\bibnamefont {Combi}, \bibfnamefont
  {M.}}, \bibinfo {author} {\bibnamefont {Keyser}, \bibfnamefont {J.~D.}},
  \bibinfo {author} {\bibnamefont {Dhooghe}, \bibfnamefont {F.}}, \bibinfo
  {author} {\bibnamefont {Eberhardt}, \bibfnamefont {P.}}, \bibinfo {author}
  {\bibnamefont {Fiethe}, \bibfnamefont {B.}}, \bibinfo {author} {\bibnamefont
  {Fuselier}, \bibfnamefont {S.~A.}}, \bibinfo {author} {\bibnamefont {Gasc},
  \bibfnamefont {S.}}, \bibinfo {author} {\bibnamefont {Gombosi}, \bibfnamefont
  {T.~I.}}, \bibinfo {author} {\bibnamefont {Hansen}, \bibfnamefont {K.~C.}},
  \bibinfo {author} {\bibnamefont {Hässig}, \bibfnamefont {M.}}, \bibinfo
  {author} {\bibnamefont {Jäckel}, \bibfnamefont {A.}}, \bibinfo {author}
  {\bibnamefont {Kopp}, \bibfnamefont {E.}}, \bibinfo {author} {\bibnamefont
  {Korth}, \bibfnamefont {A.}}, \bibinfo {author} {\bibnamefont {Roy},
  \bibfnamefont {L.~L.}}, \bibinfo {author} {\bibnamefont {Mall}, \bibfnamefont
  {U.}}, \bibinfo {author} {\bibnamefont {Marty}, \bibfnamefont {B.}}, \bibinfo
  {author} {\bibnamefont {Mousis}, \bibfnamefont {O.}}, \bibinfo {author}
  {\bibnamefont {Owen}, \bibfnamefont {T.}}, \bibinfo {author} {\bibnamefont
  {Rème}, \bibfnamefont {H.}}, \bibinfo {author} {\bibnamefont {Sémon},
  \bibfnamefont {T.}}, \bibinfo {author} {\bibnamefont {Tzou}, \bibfnamefont
  {C.-Y.}}, \bibinfo {author} {\bibnamefont {Waite}, \bibfnamefont {J.~H.}}, \
  and\ \bibinfo {author} {\bibnamefont {Wurz}, \bibfnamefont {P.}},\ }\bibfield
   {title} {\enquote {\bibinfo {title} {Molecular nitrogen in comet
  67p/churyumov-gerasimenko indicates a low formation temperature},}\ }\href
  {\doibase 10.1126/science.aaa6100} {\bibfield  {journal} {\bibinfo  {journal}
  {Science}\ }\textbf {\bibinfo {volume} {348}},\ \bibinfo {pages} {232--235}
  (\bibinfo {year} {2015})}\BibitemShut {NoStop}%
\bibitem [{\citenamefont {Sacchi}\ \emph {et~al.}(2013)\citenamefont {Sacchi},
  \citenamefont {Jaouen}, \citenamefont {Popescu}, \citenamefont {Gaudemer},
  \citenamefont {Tonnerre}, \citenamefont {Chiuzbaian}, \citenamefont {Hague},
  \citenamefont {Delmotte}, \citenamefont {Dubuisson}, \citenamefont {Cauchon},
  \citenamefont {Lagarde},\ and\ \citenamefont {Polack}}]{Sacchi_2013}%
  \BibitemOpen
  \bibfield  {author} {\bibinfo {author} {\bibnamefont {Sacchi}, \bibfnamefont
  {M.}}, \bibinfo {author} {\bibnamefont {Jaouen}, \bibfnamefont {N.}},
  \bibinfo {author} {\bibnamefont {Popescu}, \bibfnamefont {H.}}, \bibinfo
  {author} {\bibnamefont {Gaudemer}, \bibfnamefont {R.}}, \bibinfo {author}
  {\bibnamefont {Tonnerre}, \bibfnamefont {J.~M.}}, \bibinfo {author}
  {\bibnamefont {Chiuzbaian}, \bibfnamefont {S.~G.}}, \bibinfo {author}
  {\bibnamefont {Hague}, \bibfnamefont {C.~F.}}, \bibinfo {author}
  {\bibnamefont {Delmotte}, \bibfnamefont {A.}}, \bibinfo {author}
  {\bibnamefont {Dubuisson}, \bibfnamefont {J.~M.}}, \bibinfo {author}
  {\bibnamefont {Cauchon}, \bibfnamefont {G.}}, \bibinfo {author} {\bibnamefont
  {Lagarde}, \bibfnamefont {B.}}, \ and\ \bibinfo {author} {\bibnamefont
  {Polack}, \bibfnamefont {F.}},\ }\bibfield  {title} {\enquote {\bibinfo
  {title} {The {SEXTANTS} beamline at {SOLEIL}: a new facility for elastic,
  inelastic and coherent scattering of soft x-rays},}\ }\href {\doibase
  10.1088/1742-6596/425/7/072018} {\bibfield  {journal} {\bibinfo  {journal}
  {Journal of Physics: Conference Series}\ }\textbf {\bibinfo {volume} {425}},\
  \bibinfo {pages} {072018} (\bibinfo {year} {2013})}\BibitemShut {NoStop}%
\bibitem [{\citenamefont {Sodhi}\ and\ \citenamefont
  {Brion}(1984)}]{SODHI_1984}%
  \BibitemOpen
  \bibfield  {author} {\bibinfo {author} {\bibnamefont {Sodhi}, \bibfnamefont
  {R.~N.}}\ and\ \bibinfo {author} {\bibnamefont {Brion}, \bibfnamefont {C.}},\
  }\bibfield  {title} {\enquote {\bibinfo {title} {Reference energies for inner
  shell electron energy-loss spectroscopy},}\ }\href {\doibase
  https://doi.org/10.1016/0368-2048(84)80050-X} {\bibfield  {journal} {\bibinfo
   {journal} {Journal of Electron Spectroscopy and Related Phenomena}\ }\textbf
  {\bibinfo {volume} {34}},\ \bibinfo {pages} {363--372} (\bibinfo {year}
  {1984})}\BibitemShut {NoStop}%
\bibitem [{\citenamefont {Straub}\ \emph {et~al.}(1996)\citenamefont {Straub},
  \citenamefont {Renault}, \citenamefont {Lindsay}, \citenamefont {Smith},\
  and\ \citenamefont {Stebbings}}]{straub_absolute_1996}%
  \BibitemOpen
  \bibfield  {author} {\bibinfo {author} {\bibnamefont {Straub}, \bibfnamefont
  {H.~C.}}, \bibinfo {author} {\bibnamefont {Renault}, \bibfnamefont {P.}},
  \bibinfo {author} {\bibnamefont {Lindsay}, \bibfnamefont {B.~G.}}, \bibinfo
  {author} {\bibnamefont {Smith}, \bibfnamefont {K.~A.}}, \ and\ \bibinfo
  {author} {\bibnamefont {Stebbings}, \bibfnamefont {R.~F.}},\ }\bibfield
  {title} {\enquote {\bibinfo {title} {Absolute partial cross sections for
  electron-impact ionization of {H} 2 , {N} 2 , and {O} 2 from threshold to
  1000 {eV}},}\ }\href {\doibase 10.1103/PhysRevA.54.2146} {\bibfield
  {journal} {\bibinfo  {journal} {Physical Review A}\ }\textbf {\bibinfo
  {volume} {54}},\ \bibinfo {pages} {2146--2153} (\bibinfo {year}
  {1996})}\BibitemShut {NoStop}%
\bibitem [{\citenamefont {Tian}\ and\ \citenamefont
  {Vidal}(1998)}]{tian_cross_1998}%
  \BibitemOpen
  \bibfield  {author} {\bibinfo {author} {\bibnamefont {Tian}, \bibfnamefont
  {C.}}\ and\ \bibinfo {author} {\bibnamefont {Vidal}, \bibfnamefont {C.~R.}},\
  }\bibfield  {title} {\enquote {\bibinfo {title} {Cross sections of the
  electron impact dissociative ionization of {CO}, ch4 and c2h2},}\ }\href
  {\doibase 10.1088/0953-4075/31/4/031} {\bibfield  {journal} {\bibinfo
  {journal} {Journal of Physics B: Atomic, Molecular and Optical Physics}\
  }\textbf {\bibinfo {volume} {31}},\ \bibinfo {pages} {895--909} (\bibinfo
  {year} {1998})}\BibitemShut {NoStop}%
\bibitem [{\citenamefont {Valkealahti}, \citenamefont {Schou},\ and\
  \citenamefont {Nieminen}(1989)}]{valkealahti_energy_1989}%
  \BibitemOpen
  \bibfield  {author} {\bibinfo {author} {\bibnamefont {Valkealahti},
  \bibfnamefont {S.}}, \bibinfo {author} {\bibnamefont {Schou}, \bibfnamefont
  {J.}}, \ and\ \bibinfo {author} {\bibnamefont {Nieminen}, \bibfnamefont
  {R.~M.}},\ }\bibfield  {title} {\enquote {\bibinfo {title} {Energy deposition
  of kev electrons in light elements},}\ }\href {\doibase 10.1063/1.342839}
  {\bibfield  {journal} {\bibinfo  {journal} {Journal of Applied Physics}\
  }\textbf {\bibinfo {volume} {65}},\ \bibinfo {pages} {2258--2266} (\bibinfo
  {year} {1989})},\ \Eprint
  {http://arxiv.org/abs/https://doi.org/10.1063/1.342839}
  {https://doi.org/10.1063/1.342839} \BibitemShut {NoStop}%
\bibitem [{\citenamefont {Walters}\ and\ \citenamefont
  {Bhalla}(1971)}]{Walters_1971}%
  \BibitemOpen
  \bibfield  {author} {\bibinfo {author} {\bibnamefont {Walters}, \bibfnamefont
  {D.~L.}}\ and\ \bibinfo {author} {\bibnamefont {Bhalla}, \bibfnamefont
  {C.~P.}},\ }\bibfield  {title} {\enquote {\bibinfo {title} {Nonrelativistic
  auger rates, x-ray rates, and fluorescence yields for the $k$ shell},}\
  }\href {\doibase 10.1103/PhysRevA.3.1919} {\bibfield  {journal} {\bibinfo
  {journal} {Phys. Rev. A}\ }\textbf {\bibinfo {volume} {3}},\ \bibinfo {pages}
  {1919--1927} (\bibinfo {year} {1971})}\BibitemShut {NoStop}%
\bibitem [{\citenamefont {Öberg}\ \emph {et~al.}(2005)\citenamefont {Öberg},
  \citenamefont {van Broekhuizen}, \citenamefont {Fraser}, \citenamefont
  {Bisschop}, \citenamefont {van Dishoeck},\ and\ \citenamefont
  {Schlemmer}}]{oberg_competition_2005}%
  \BibitemOpen
  \bibfield  {author} {\bibinfo {author} {\bibnamefont {Öberg}, \bibfnamefont
  {K.~I.}}, \bibinfo {author} {\bibnamefont {van Broekhuizen}, \bibfnamefont
  {F.}}, \bibinfo {author} {\bibnamefont {Fraser}, \bibfnamefont {H.~J.}},
  \bibinfo {author} {\bibnamefont {Bisschop}, \bibfnamefont {S.~E.}}, \bibinfo
  {author} {\bibnamefont {van Dishoeck}, \bibfnamefont {E.~F.}}, \ and\
  \bibinfo {author} {\bibnamefont {Schlemmer}, \bibfnamefont {S.}},\ }\bibfield
   {title} {\enquote {\bibinfo {title} {Competition between {CO} and {N}
  $_{\textrm{2}}$ {Desorption} from {Interstellar} {Ices}},}\ }\href {\doibase
  10.1086/428901} {\bibfield  {journal} {\bibinfo  {journal} {The Astrophysical
  Journal}\ }\textbf {\bibinfo {volume} {621}},\ \bibinfo {pages} {L33--L36}
  (\bibinfo {year} {2005})}\BibitemShut {NoStop}%
\bibitem [{\citenamefont {Öberg}, \citenamefont {van Dishoeck},\ and\
  \citenamefont {Linnartz}(2009)}]{oberg_photodesorption_2009}%
  \BibitemOpen
  \bibfield  {author} {\bibinfo {author} {\bibnamefont {Öberg}, \bibfnamefont
  {K.~I.}}, \bibinfo {author} {\bibnamefont {van Dishoeck}, \bibfnamefont
  {E.~F.}}, \ and\ \bibinfo {author} {\bibnamefont {Linnartz}, \bibfnamefont
  {H.}},\ }\bibfield  {title} {\enquote {\bibinfo {title} {Photodesorption of
  ices {I}: {CO}, {N}2, and {CO}2},}\ }\href {\doibase
  10.1051/0004-6361/200810207} {\bibfield  {journal} {\bibinfo  {journal}
  {Astronomy \& Astrophysics}\ }\textbf {\bibinfo {volume} {496}},\ \bibinfo
  {pages} {281--293} (\bibinfo {year} {2009})}\BibitemShut {NoStop}%
\bibitem [{\citenamefont {Öberg}\ \emph {et~al.}(2007)\citenamefont {Öberg},
  \citenamefont {Fuchs}, \citenamefont {Awad}, \citenamefont {Fraser},
  \citenamefont {Schlemmer}, \citenamefont {van Dishoeck},\ and\ \citenamefont
  {Linnartz}}]{_berg_2007}%
  \BibitemOpen
  \bibfield  {author} {\bibinfo {author} {\bibnamefont {Öberg}, \bibfnamefont
  {K.~I.}}, \bibinfo {author} {\bibnamefont {Fuchs}, \bibfnamefont {G.~W.}},
  \bibinfo {author} {\bibnamefont {Awad}, \bibfnamefont {Z.}}, \bibinfo
  {author} {\bibnamefont {Fraser}, \bibfnamefont {H.~J.}}, \bibinfo {author}
  {\bibnamefont {Schlemmer}, \bibfnamefont {S.}}, \bibinfo {author}
  {\bibnamefont {van Dishoeck}, \bibfnamefont {E.~F.}}, \ and\ \bibinfo
  {author} {\bibnamefont {Linnartz}, \bibfnamefont {H.}},\ }\bibfield  {title}
  {\enquote {\bibinfo {title} {Photodesorption of {CO} ice},}\ }\href {\doibase
  10.1086/519281} {\bibfield  {journal} {\bibinfo  {journal} {The Astrophysical
  Journal}\ }\textbf {\bibinfo {volume} {662}},\ \bibinfo {pages} {L23--L26}
  (\bibinfo {year} {2007})}\BibitemShut {NoStop}%
\end{thebibliography}%

\end{document}